\documentclass[]{article}
\usepackage{proceed2e}

\usepackage{url,mathrsfs}
\usepackage{amsmath,wrapfig,color,bigfoot,times}
\usepackage{amsfonts}
\usepackage{graphicx}
\usepackage{subfigure}
\newtheorem{theorem}{Theorem}
\newtheorem{lemma}{Lemma}

\title{\vspace{-0.15in}Improved Densification of One Permutation Hashing\vspace{-0.15in}}

\author{
{\bf Anshumali Shrivastava}\\
Department of Computer Science\\
Computing and Information Science\\
Cornell University\\
Ithaca, NY 14853, USA\\
\texttt{anshu@cs.cornell.edu}
\And
{\bf Ping Li} \\
Department of Statistics and Biostatistics\\
Department of Computer Science \\
Rutgers University\\
Piscataway, NJ 08854, USA \\
\texttt{pingli@stat.rutgers.edu}
}

\begin{document}

\maketitle

\begin{abstract}\vspace{-0.2in}
The existing work on densification of one permutation hashing~\cite{Proc:OneHashLSH_ICML14} reduces the query processing cost of the $(K,L)$-parameterized Locality Sensitive Hashing (LSH) algorithm with minwise hashing, from $O(dKL)$  to merely $O(d + KL)$,  where $d$ is the  number of nonzeros of the data vector, $K$ is the number of hashes in each hash table, and $L$ is the number of hash tables. While that is a substantial improvement, our analysis reveals that the existing densification scheme in~\cite{Proc:OneHashLSH_ICML14}  is sub-optimal.  In particular, there is no enough randomness in that procedure, which affects its accuracy on very sparse datasets.

\noindent In this paper, we provide a new densification procedure which is provably better than the existing scheme~\cite{Proc:OneHashLSH_ICML14}. This improvement is more significant for very sparse datasets which are common over the web. The improved technique has the same cost of $O(d + KL)$ for query processing, thereby making it strictly preferable over the existing  procedure. Experimental evaluations on public datasets, in the task of hashing based near neighbor search,  support our theoretical findings.
\end{abstract}

\vspace{-0.25in}
\section{Introduction}
\vspace{-0.1in}
Binary representations are common for high dimensional sparse data over the web~\cite{Sibyl,GoogleBlog,Proc:Weinberger_ICML2009,Report:TeraLarning11}, especially for text data represented by high-order $n$-grams~\cite{Proc:Broder,Proc:Fetterly_WWW03}. Binary vectors can also be equivalently viewed as sets, over the universe of all the features, containing only  locations of the non-zero entries. Given two sets $S_1$, $S_2 \subseteq \Omega = \{1,2,...,D\}$, a popular measure of similarity between sets (or binary vectors) is the {\em resemblance} $R$, defined as
\begin{equation}
 R = \frac{|S_1 \cap S_2|}{| S_1 \cup S_2|} = \frac{a}{f_1+f_2 -a},
\end{equation}
where $f_1 = |S_1|$, $f_2 = |S_2|$, and $a = |S_1 \cap S_2|$.

It is well-known that minwise hashing belongs to the {\em Locality Sensitive Hashing (LSH)} family~\cite{Proc:Broder_STOC98,Proc:Charikar}. The method applies a random permutation $\pi:\Omega \rightarrow \Omega$, on the given set $S$, and stores the minimum value after the permutation mapping. Formally,
 \begin{equation}h_{\pi}(S) = \min(\pi(S)).\end{equation} Given sets $S_1$ and $S_2$, it can be shown by elementary probability arguments that
\begin{equation}
\label{eq:minhash}
Pr({h_{\pi}(S_1) = h_{\pi}(S_2)) =  \frac{|S_1 \cap S_2|}{| S_1 \cup S_2|}} = R.
\end{equation}
The probability of collision (equality of hash values), under minwise hashing, is equal to the similarity of interest $R$. This property, also known as the {\em LSH property}~\cite{Proc:Indyk_STOC98,Proc:Charikar}, makes minwise hash functions $h_{\pi}$ suitable for creating hash buckets, which leads to sublinear algorithms for similarity search. Because of this same  LSH property, minwise hashing is a popular indexing technique for a variety of large-scale data processing applications, which include duplicate detection~\cite{Proc:Broder,Proc:Henzinger_SIGIR06}, all-pair similarity~\cite{Proc:Bayardo_WWW07}, fast linear learning~\cite{Proc:HashLearning_NIPS11}, temporal correlation~\cite{Proc:Chien_WWW05}, 3-way similarity \& retrieval~\cite{Proc:Li_Konig_NIPS10,Proc:Shrivastava_NIPS13}, graph algorithms~\cite{Proc:Buehrer_WSDM08,Proc:Chierichetti_KDD09,Proc:Najork_WSDM09}, and  more.

Querying with a standard $(K,L)$-parameterized LSH algorithm~\cite{Proc:Indyk_STOC98}, for fast similarity search, requires computing $K \times L$ min-hash values per query, where $K$ is the number of hashes in each hash table and $L$ is the number of hash tables.  In theory, the value of $K L$ grows with the data size~\cite{Proc:Indyk_STOC98}. In practice, typically, this number ranges from a few hundreds to a few thousands.  Thus, processing a single query, for near-neighbor search, requires  evaluating hundreds or thousands of independent permutations $\pi$ (or cheaper universal approximations to permutations~\cite{Proc:Carter_STOC77,Proc:Nisan_STOC90,Proc:Mitzenmacher_SODA08}) over the given data vector.  If $d$ denotes the number of non-zeros in the query vector, then the query preprocessing cost is $O(dKL)$ which is also the bottleneck step in the LSH algorithm~\cite{Proc:Indyk_STOC98}. Query time (latency) is crucial in many  user-facing applications, such as search.

Linear learning with $b$-bit minwise hashing~\cite{Proc:HashLearning_NIPS11}, requires multiple evaluations (say $k$) of  $h_{\pi}$ for a given data vector.  Computing $k$ different min-hashes of the test data costs $O(dk)$, while after processing, classifying this data vector (with SVM or logistic regression) only requires a single inner product with the weight vector which is $O(k)$. Again, the bottleneck step during testing prediction is the evaluation of $k$ min-hashes. Testing time directly translates into the latency of on-line classification systems.

The idea of storing $k$ contiguous minimum values after one single permutation~\cite{Proc:Broder,Proc:Li_Church_EMNLP,Proc:Li_Church_Hastie_NIPS06}  leads to hash values which do not satisfy the LSH property because the hashes are not properly aligned.  The estimators are also not linear, and therefore they do not lead to feature representation for linear learning with resemblance. This is a serious limitation.

Recently it was shown that a ``rotation" technique~\cite{Proc:OneHashLSH_ICML14} for densifying  sparse sketches  from one permutation hashing~\cite{Proc:Li_Owen_Zhang_NIPS12} solves the problem of costly processing with minwise hashing (See Sec.~\ref{sec:background}). The scheme only requires a single permutation and generates $k$  different hash values, satisfying the LSH property (i.e., Eq.(\ref{eq:minhash})),  in linear time $O(d + k)$,  thereby reducing a factor  $d$ in the processing cost compared to the original minwise hashing.

{\bf Our Contributions:} In this paper, we argue that the existing densification scheme~\cite{Proc:OneHashLSH_ICML14} is not the optimal way of densifying the sparse sketches of one permutation hashing at the given processing cost. In particular, we provide a provably better densification scheme for generating $k$ hashes with the same processing cost of $O(d +k)$. Our contributions can be summarized as follows.
\begin{itemize}\vspace{-0.1in}
\item Our detailed variance analysis of the hashes obtained from the existing densification scheme~\cite{Proc:OneHashLSH_ICML14} reveals that there is no enough randomness in that procedure which leads to high variance in very sparse datasets.\vspace{-0.05in}
\item We provide a new densification scheme for one permutation hashing with provably smaller variance than the scheme in~\cite{Proc:OneHashLSH_ICML14}. The improvement becomes more significant for very sparse datasets which are common in practice.  The improved scheme retains the computational complexity of $O(d+k)$ for computing $k$ different hash evaluations of a given vector.\vspace{-0.05in}
\item We provide experimental evidences on publicly available datasets, which demonstrate the superiority of the improved densification procedure over the existing scheme, in the task of resemblance estimation and as well as the task of near neighbor retrieval with LSH.
\end{itemize}

\vspace{-0.15in}
\section{Background}\label{sec:background}
\vspace{-0.05in}
\subsection{One Permutation Hashing}
\begin{figure}[h]
\begin{center}
\mbox{
\includegraphics[width=3.2in]{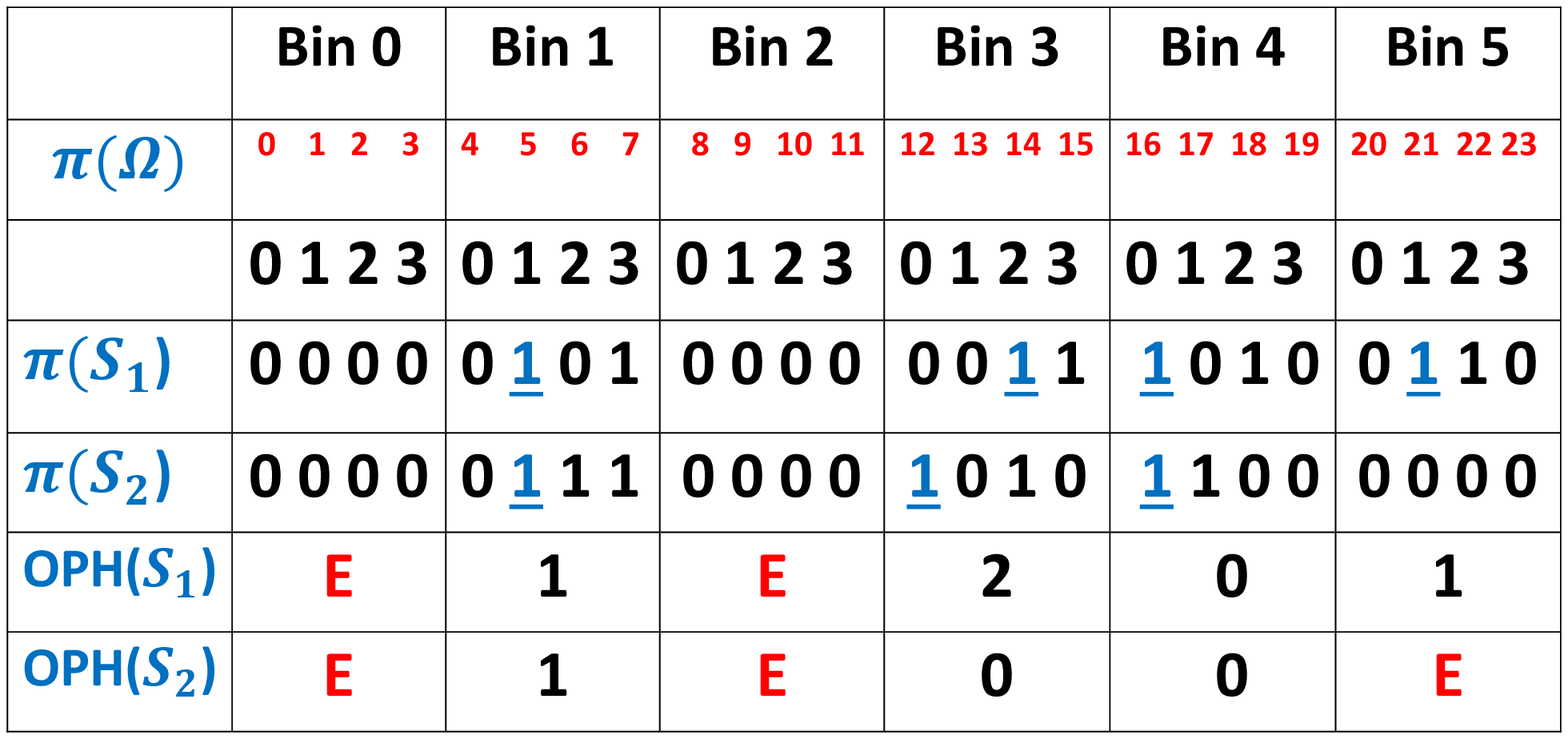} }
\end{center}
\vspace{-0.35in} \caption{One permutation hashes~\cite{Proc:Li_Owen_Zhang_NIPS12} for vectors $S_1$ and $S_2$ using a single permutation $\pi$. For bins not containing any non-zeros, we use special symbol ``E".}\label{fig:1phash}
\end{figure}

\vspace{-0.05in}
As illustrated in Figure~\ref{fig:1phash}, instead of conducting $k$ independent permutations, {\em one permutation hashing}~\cite{Proc:Li_Owen_Zhang_NIPS12} uses only one permutation and  partitions the (permuted) feature space into  $k$ bins. In other words,  a single permutation $\pi$ is used to first shuffle the given binary vector, and then the shuffled vector is binned into $k$ evenly spaced bins. The $k$ minimums, computed for each bin separately, are the $k$ different hash values. Obviously, empty bins are possible.

For example, in Figure~\ref{fig:1phash}, $\pi(S_1)$ and $\pi(S_2)$ denote the state of the binary vectors $S_1$ and $S_2$ after applying permutation $\pi$. These shuffled vectors are then divided into $6$ bins of length $4$ each. We start the numbering from 0.  We  look into each bin and store the corresponding minimum non-zero index.  For bins not containing any non-zeros, we use a special symbol ``E'' to denote empty bins. We also denote
\begin{equation}
M_j(\pi(S)) = \left\{\pi(S) \cap \left[\frac{Dj}{k},\frac{D(j+1)}{k}\right)\right\}
\end{equation}
We assume for the rest of the paper that $D$ is divisible by $k$, otherwise we can always pad extra dummy features. We define $\underset{j}{OPH}$ (``OPH'' for one permutation hashing) as
\begin{align}
\underset{j}{OPH}(\pi(S))  = \begin{cases}E, & \hspace{-0.85in}\mbox{if }\pi(S) \cap \left[\frac{Dj}{k},\frac{D(j+1)}{k}\right) = \phi \\ M_j(\pi(S)) \hspace{-0.1in}\mod{\frac{D}{k}}, & \mbox{otherwise}\end{cases}
\end{align}
i.e., $\underset{j}{OPH}(\pi(S))$ denotes the minimum value in Bin $j$, under permutation mapping $\pi$, as shown in the example in Figure~\ref{fig:1phash}. If this intersection is null, i.e., $\pi(S) \cap \left[\frac{Dj}{k},\frac{D(j+1)}{k}\right) = \phi$, then $\underset{j}{OPH}(\pi(S)) =E$.

Consider the events of ``simultaneously empty bin"  $I_{emp}^j =1$ and ``simultaneously non-empty bin" $I_{emp}^j =0$, between given vectors $S_1$ and $S_2$, defined as:
\begin{align}
I_{emp}^j= \begin{cases} 1, &\mbox{if $\underset{j}{OPH}(\pi(S_1)) = \underset{j}{OPH}(\pi(S_2)) = E$}\\ 0 &\mbox{otherwise}\end{cases}
\end{align}\vspace{-0.25in}

 Simultaneously empty bins are only defined with respect to two sets $S_1$ and $S_2$.  In Figure~\ref{fig:1phash}, $I_{emp}^0 = 1$ and $I_{emp}^2 = 1$, while $I_{emp}^1 =  I_{emp}^3 = I_{emp}^4 =  I_{emp}^5 =  0$. Bin 5 is only empty for $S_2$ and not for $S_1$, so $I_{emp}^5 =  0$.

Given a bin number $j$, if it is not simultaneously empty ($I_{emp}^j =0$)  for both the vectors $S_1$ and $S_2$, \cite{Proc:Li_Owen_Zhang_NIPS12} showed
\begin{equation}
\label{eq:1pprob}
Pr\left(\underset{j}{OPH}(\pi(S_1)) = \underset{j}{OPH}(\pi(S_2)) \bigg| I_{emp}^j = 0 \right) =  R
\end{equation}

On the other hand, when $I_{emp}^j =1$, no such guarantee exists. When $I_{emp}^j =1$ collision does not have enough information about the similarity $R$.  Since the event $I_{emp}^j =1$ can only be determined  given the two vectors $S_1$ and $S_2$ and the materialization of $\pi$, one permutation hashing cannot be directly used for indexing, especially when the data are very sparse.  In particular, $\underset{j}{OPH}(\pi(S))$ does not lead to a valid LSH hash function because of the coupled event $I_{emp}^j =1$ in (\ref{eq:1pprob}). The simple strategy of ignoring empty bins leads to biased estimators of resemblance and shows poor performance~\cite{Proc:OneHashLSH_ICML14}. Because of this same reason, one permutation hashing cannot be directly used to extract random features for linear learning with resemblance kernel.

\vspace{-0.05in}
\subsection{Densifying One Permutation Hashing for Indexing and Linear Learning}
\label{sec:old_densi}
\vspace{-0.05in}

\begin{figure}[h!]
\begin{center}
\mbox{
\includegraphics[width=3.2in]{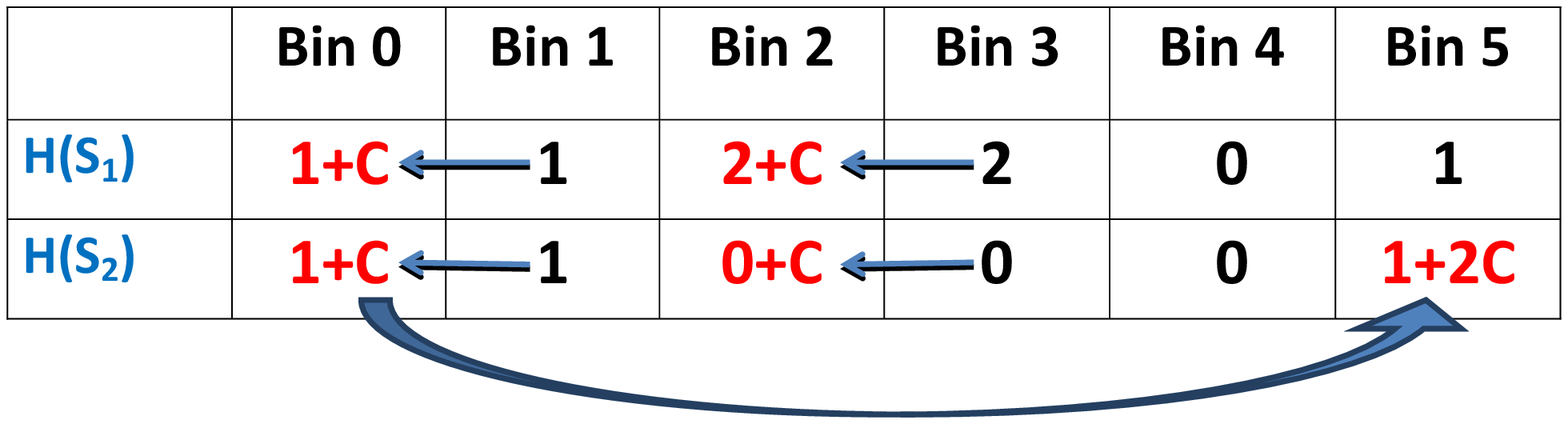} }
\end{center}
\vspace{-0.25in} \caption{Densification  by ``rotation'' for filling empty bins generated from one permutation hashing~\cite{Proc:OneHashLSH_ICML14}. Every empty bin is assigned the value of the closest non-empty bin, towards right (circular), with an offset $C$. For the configuration shown in Figure~\ref{fig:1phash}, the above figure shows the  new assigned values (in red) of empty bins after densification.}\label{fig:1protation}
\end{figure}

\cite{Proc:OneHashLSH_ICML14} proposed a ``rotation" scheme that assigns new values to all the empty bins, generated from one permutation hashing, in an unbiased fashion. The rotation scheme for filling the empty bins from Figure~\ref{fig:1phash} is shown in Figure~\ref{fig:1protation}.  The idea is that for every empty bin, the scheme borrows the value of the closest non-empty bin in the clockwise direction (circular right hand side) added with offset  $C$.

Given the configuration in Figure~\ref{fig:1phash}, for Bin 2 corresponding to $S_1$, we borrow the value 2 from Bin 3 along with an additional offset of $C$. Interesting is the case of Bin 5 for $S_2$, the circular right is Bin 0 which was empty. Bin 0 borrows from Bin 1 acquiring value $1+C$, Bin 5 borrows this value with another offset $C$. The value of Bin 5 finally becomes $1+2C$. The value of $C=\frac{D}{k}+1$ enforces proper alignment and ensures no unexpected collisions. Without this offset $C$, Bin 5, which was not simultaneously empty, after reassignment, will have value 1 for both $S_1$ and $S_2$. This would be an error as initially there was no collision (note $I_{emp}^5 = 0$). Multiplication by the distance of the non-empty bin, from where the value was borrowed, ensures that the new values of simultaneous empty bins ($I_{emp}^j =1$), at any location $j$ for $S_1$ and $S_2$,  never match if their new values come from different bin numbers.

Formally the  hashing scheme with ``rotation'', denoted by $\mathcal{H}$, is defined as:
\begin{align}\label{eqn_H}
&\mathcal{H}_j(S) =\left\{\begin{array}{ll}
\underset{j}{OPH}(\pi(S)) &\hspace{-0.6in}\text {if }\underset{j}{OPH}(\pi(S))   \neq E\\\\
\underset{(j+t)\ \text{mod}\ k}{OPH}(\pi(S)) + t C &\text{otherwise}
\end{array}
\right.\\
&t = \min z,\ \ \ s.t.\ \ \underset{(j+z)\ \text{mod}\ k}{OPH}(\pi(S))\neq E
\end{align}
Here  $C = \frac{D}{k}+1$ is a constant.

This densification scheme ensures that whenever $I_{emp}^j = 0$, i.e., Bin $j$ is simultaneously empty for any two $S_1$  and $S_2$ under considerations, the newly assigned value mimics the collision probability of the nearest simultaneously non-empty bin towards right (circular) hand side making the final collision probability equal to $R$, irrespective of whether $I_{emp}^j=0$ or $I_{emp}^j=1$.  \cite{Proc:OneHashLSH_ICML14} proved this fact as a theorem.
\begin{theorem}\label{thm_H_Pr_old}\cite{Proc:OneHashLSH_ICML14}
\begin{align}
\label{eq:coll_prob_old}
\mathbf{Pr}\left(\mathcal{H}_j(S_1) = \mathcal{H}_j(S_2)\right) = R
\end{align}
\end{theorem}
Theorem~\ref{thm_H_Pr_old} implies that $\mathcal{H}$ satisfies the LSH property and hence it is suitable for indexing based sublinear similarity search. Generating $KL$ different hash values of $\mathcal{H}$ only requires $O(d + KL)$, which saves a factor of $d$ in the query processing cost compared to the cost of $O(dKL)$ with traditional minwise hashing. For fast linear learning~\cite{Proc:HashLearning_NIPS11} with $k$ different hash values the new scheme only needs $O(d + k)$ testing (or prediction) time compared to  standard $b$-bit minwise hashing which requires $O(dk)$ time for testing.

\vspace{-0.1in}
\section{Variance Analysis of Existing Scheme}\label{sec:var_old}
\vspace{-0.05in}

We first provide the variance analysis of the existing scheme~\cite{Proc:OneHashLSH_ICML14}. Theorem~\ref{thm_H_Pr_old} leads to an unbiased estimator of $R$ between $S_1$ and $S_2$ defined as:
\begin{equation}
\hat{R} = \frac{1}{k}\sum_{j=0}^{k-1} \mathbf{1}\{\mathcal{H}_{j}(S_1) = \mathcal{H}_{j}(S_2)\}.
\end{equation}
Denote the number of simultaneously empty bins by
\begin{align}
N_{emp} = \sum_{j =0}^{k-1}\mathbf{1}\{I_{emp}^j =1\},
\end{align}
where $\mathbf{1}$ is the indicator function. We  partition the event $\left(\mathcal{H}_j(S_1) = \mathcal{H}_j(S_2)\right)$ into two cases depending on $I_{emp}^j$. Let $M_j^N$ ({\em Non-empty Match at $j$}) and $M_j^E$ ({\em Empty Match at $j$}) be the events defined as:
\begin{align}
M_j^N &= \mathbf{1}\{I_{emp}^j =0\ \mbox{ and }\ \mathcal{H}_j(S_1) = \mathcal{H}_j(S_2)\} \\
M_j^E &= \mathbf{1}\{I_{emp}^j =1\ \mbox{ and }\ \mathcal{H}_j(S_1) = \mathcal{H}_j(S_2)\}
\end{align}
Note that,
$M_j^N = 1 \implies M_j^E = 0$ and $M_j^E = 1 \implies M_j^N = 0.$
This combined with Theorem~\ref{thm_H_Pr_old} implies,
\begin{align}
\mathbb{E}(M_j^N| I_{emp}^j = 0) &= \mathbb{E}(M_j^E | I_{emp}^j =1) \notag\\
&=\mathbb{E}( M_j^E + M_j^N ) = R  \ \ \forall j
\end{align}
It is not difficult to show that,
\begin{align}\notag
\mathbb{E}\left(M_j^NM_i^N\big| i \ne j, I_{emp}^j = 0 \ \text{ and}  \ I_{emp}^i = 0\right) = R\tilde{R},
\end{align}
where $\tilde{R} = \frac{a-1}{f1 + f2 - a -1}$.  Using these new events, we have
\begin{align}
\hat{R} = \frac{1}{k}\sum_{j=0}^{k-1} \left[ M_j^E + M_j^N \right]
\end{align}
We are interested in computing
\begin{align}
Var(\hat{R}) = \mathbb{E}\left(\left(\frac{1}{k}\sum_{j=0}^{k-1} \left[ M_j^E + M_j^N \right]\right )^2\right) - R^2
\end{align}

For notational convenience we will use $m$ to denote the event $k - N_{emp}=m$, i.e., the expression $\mathbb{E}( . | m)$ means $\mathbb{E}( . |k - N_{emp} = m).$ To simplify the analysis, we will first compute the conditional expectation
\begin{align}
f(m) = \mathbb{E}\left(\left(\frac{1}{k}\sum_{j=0}^{k-1} \left[ M_j^E + M_j^N \right]\right )^2\bigg| \ m\right)
\end{align}
 By expansion  and  linearity of expectation, we obtain
\begin{align}
k^2f(m) =  \mathbb{E}\left[\sum_{i \ne j}M_i^N M_j^N\bigg|  m\right] + \mathbb{E}\left[\sum_{i \ne j}M_i^N M_j^E\bigg|m\right]\notag \\
    + \mathbb{E}\left[\sum_{i \ne j}M^E_i M^E_j\bigg|m\right]  + \mathbb{E}\left[\sum_{i =1}^{k}\left[(M_j^N)^2 +(M_j^E)^2\right]\bigg| m\right]\notag
\end{align}
$M_j^N = (M_j^N )^2 $ and $M_j^E = (M_j^E)^2$ as they are indicator functions and can only take values 0 and 1. Hence,
\begin{align}
\label{eq:exp_square}
 \mathbb{E}\left[\sum_{j =0}^{k-1}\left[(M_j^N)^2 +(M_j^E)^2\right]\bigg| m\right]= kR
\end{align}
The values of the remaining three terms are given by the following 3 Lemmas; See the proofs in the Appendix.
\begin{lemma}
\label{lem:nonemptynonempty}
\begin{align}
\mathbb{E}\left[\sum_{i \ne j}M_i^N M_j^N\bigg|  m\right]  =  m(m-1)R\tilde{R}
\end{align}
\end{lemma}
\begin{lemma}
\label{lem:emptyandnonempty}
\begin{align}
\mathbb{E}\left[\sum_{i \ne j}M_i^N M_j^E\bigg|m\right]  =  2m(k-m)\left[\frac{R}{m} + \frac{(m-1)R\tilde{R}}{m}\right]
\end{align}
\end{lemma}

\begin{lemma}
\label{lem:emptemptyold}
\begin{align}
 \mathbb{E}\left[\sum_{i \ne j}M^E_i M^E_j\bigg|m\right]  &=  (k-m)(k-m-1) \notag\\
&\times \left[\frac{2R}{m+1} + \frac{(m-1)R\tilde{R}}{m+1}\right]
\end{align}
\end{lemma}

Combining the expressions from the above 3 Lemmas and Eq.(\ref{eq:exp_square}), we can compute $f(m)$. Taking a further expectation over values of $m$ to remove the conditional dependency, the  variance of $\hat{R}$ can be shown in the next Theorem.
\begin{theorem}
\label{theo:var_old}
\begin{align}
&Var(\hat{R}) =\frac{R}{k} + A\frac{R}{k} + B\frac{R\tilde{R}}{k} - R^2 \\
&A = 2\mathbb{E}\left[\frac{N_{emp}}{k - N_{emp} +1}\right]\notag\\
&B =(k+ 1)\mathbb{E}\left[\frac{k - N_{emp}-1}{k - N_{emp} +1}\right]\notag
\end{align}
The theoretical values of $A$ and $B$ can be computed using the probability of the event $Pr(N_{emp} = i)$, denoted by  $P_i$,  which is given by Theorem 3 in~\cite{Proc:Li_Owen_Zhang_NIPS12}.
\begin{align}\notag
P_i = \sum_{s=0}^{k-i}\frac{(-1)^s k!}{i!s!(k-i-s)!}\prod_{t =0}^{f_1+f_2-a-1}\frac{D\left(1 - \frac{i+s}{k}\right) -t}{D-t}
\end{align}
\end{theorem}

\vspace{-0.1in}
\section{Intuition for the Improved Scheme}\label{sec:intut}
\vspace{-0.2in}

 \begin{figure}[h!]
\begin{center}
\mbox{
\includegraphics[width=3.2in]{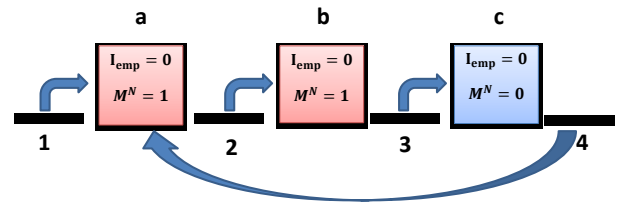} }
\end{center}
\vspace{-0.25in} \caption{Illustration of the existing  densification scheme~\cite{Proc:OneHashLSH_ICML14}. The 3 boxes indicate 3 simultaneously non-empty bins. Any simultaneously empty bin has 4  possible positions shown by blank spaces. Arrow indicates the choice of simultaneous non-empty  bins picked by  simultaneously empty bins occurring in the corresponding positions. A  simultaneously empty bin occurring in position 3 uses the information from Bin $c$. The randomness is in the position number of these bins which depends on $\pi$.}\label{fig:cartoon_old}
\end{figure}

Consider a situation in Figure~\ref{fig:cartoon_old}, where there are   3  simultaneously non-empty bins ($I_{emp} = 0$) for given $S_1$ and $S_2$.  The actual position numbers of these simultaneously non-empty bins are random.  The  simultaneously empty bins ($I_{emp} = 1$) can occur  in any order in the 4  blank spaces.  The arrows in the figure show the simultaneously non-empty bins which are being picked by the simultaneously empty bins ($I_{emp} = 1$) located in the shown blank spaces. The randomness in the system is in the ordering of simultaneously empty and simultaneously non-empty bins.

Given a simultaneously non-empty Bin $t$ ($I_{emp}^t = 0$), the probability that it is picked by a given simultaneously empty Bin $i$ ($I_{emp}^i =1$) is exactly $\frac{1}{m}$. This is because the permutation $\pi$ is perfectly random and  given $m$, any ordering of $m$ simultaneously non-empty bins and $k-m$ simultaneously empty bins are equally likely. Hence, we obtain the term $\left[\frac{R}{m} + \frac{(m-1)R\tilde{R}}{m}\right]$ in Lemma~\ref{lem:emptyandnonempty}.

On the other hand, under the given scheme, the probability that  two simultaneously empty bins, $i$ and $j$,  (i.e., $I_{emp}^i =1$, $I_{emp}^j =1$),  both pick the same simultaneous non-empty Bin $t$ ($I_{emp}^t = 0$) is given by (see proof of Lemma~\ref{lem:emptemptyold})
\begin{align}
\label{eq:p}
p = \frac{2}{m+1}
\end{align}
The value of $p$ is high because there is no enough randomness in the selection procedure. Since $R \le 1$ and $R \le R\tilde{R}$, if we can reduce this probability $p$ then we reduce the value of $[pR + (1-p)R\tilde{R}]$. This directly reduces the value of $(k-m)(k-m-1) \left[\frac{2R}{m+1} + \frac{(m-1)R\tilde{R}}{m+1}\right]$ as given by Lemma~\ref{lem:emptemptyold}. The reduction scales with $N_{emp}$.

For every simultaneously empty bin, the current scheme uses the information of the closest non-empty bin in the right. Because of the symmetry in the arguments, changing the direction to left  instead of right also leads to a valid densification scheme with exactly same variance.  This is where we can infuse randomness without violating the alignment necessary for unbiased densification. We show that  randomly switching between left and right provably improves (reduces) the variance by making the sampling procedure of simultaneously non-empty bins more random.

\vspace{-0.1in}
\section{The Improved Densification Scheme}
\vspace{-0.2in}

 \begin{figure}[h]
\begin{center}
\mbox{
\includegraphics[width=3.2in]{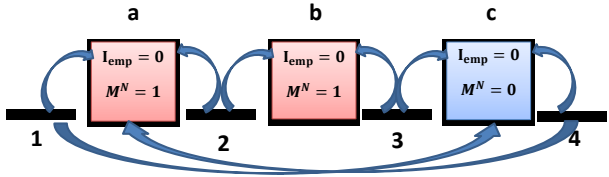} }
\end{center}
\vspace{-0.2in} \caption{Illustration of the improved densification scheme. For every simultaneously empty bin, in the blank position, instead of always choosing the simultaneously non-empty bin from right, the new scheme randomly chooses to go either left or right. A simultaneously empty bin occurring at position 2  uniformly chooses among  Bin $a$ or  Bin $b$.}\label{fig:cartoon_new}
\end{figure}

Our proposal is explained in Figure~\ref{fig:cartoon_new}. Instead of using the value of the closest non-empty bin from the right (circular), we will choose to go either left or right with probability $\frac{1}{2}$. This adds more randomness in the selection procedure.

\begin{figure}[h!]
\begin{center}
\mbox{
\includegraphics[width=3.2in]{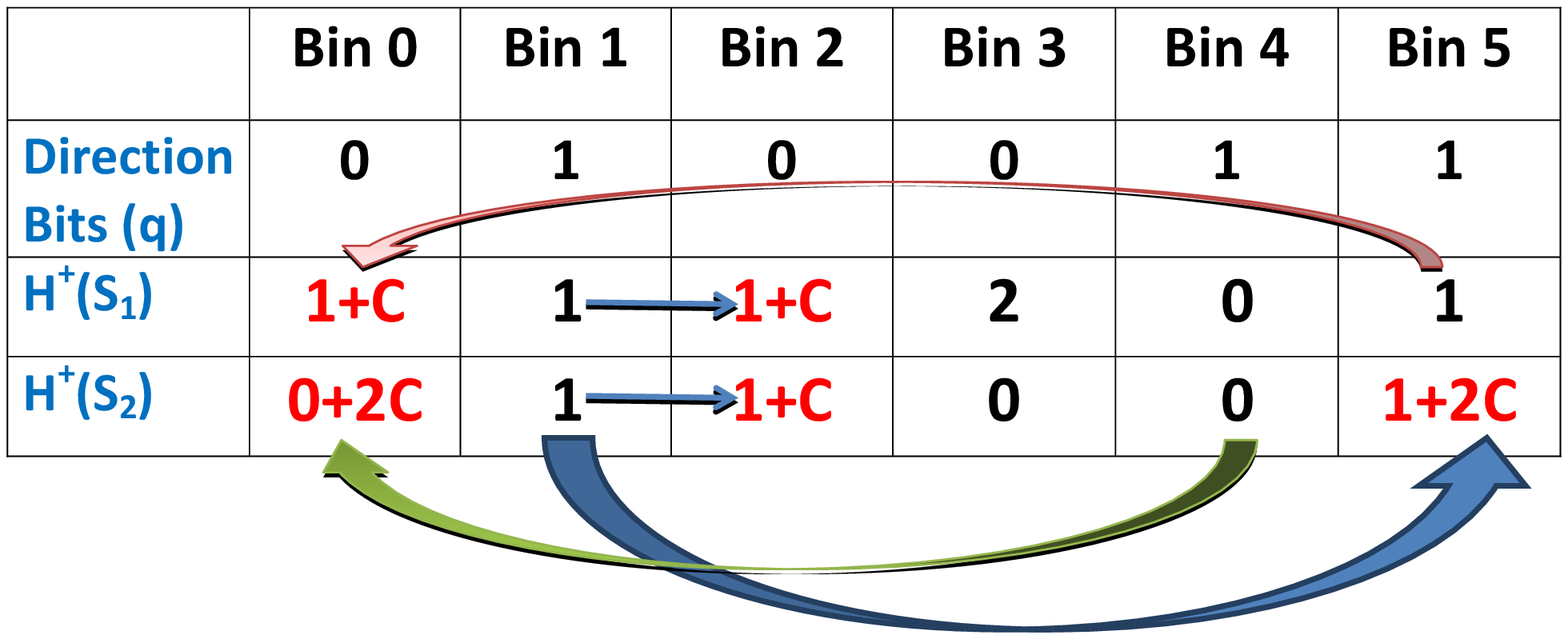} }
\end{center}
\vspace{-0.2in} \caption{  Assigned values (in red) of empty bins from Figure~\ref{fig:1phash} using the improved densification procedure. Every empty Bin $i$ uses the value of the closest non-empty bin, towards circular left or circular right depending on the random direction bit $q_i$, with offset $C$.}\label{fig:1protationImp}
\end{figure}

In the new  scheme, we only need to store $1$ random bit for each bin, which decides the direction (circular left or circular right) to proceed for finding the closest non-empty bin.  The new assignment of the empty bins from Figure~\ref{fig:1phash} is shown in Figure~\ref{fig:1protationImp}. Every bin number $i$ has an i.i.d. Bernoulli random variable $q_i$ (1 bit) associated with it.  If Bin $i$ is empty, we check the value of $q_i$. If $q_i =1$, we move circular right to find the closest non-empty bin and use its value. In case when $q=0$, we move circular left.

For $S_1$, we have $q_0 = 0$ for empty Bin 0, we therefore move circular left and borrow value from Bin 5 with offset $C$ making the final value $1+C$.  Similarly for empty Bin 2 we have $q_2 = 0$ and we use the value of Bin 1 (circular left) added with $C$.  For $S_2$ and Bin 0, we have $q_0 = 0$ and the next circular left bin is Bin 5 which is empty so we continue and borrow value from Bin 4, which is 0, with offset $2C$. It is a factor of 2 because we traveled 2 bins to locate the first non-empty bin. For Bin 2, again $q_2 = 0$ and the closest circular left non-empty bin is Bin 1, at distance 1, so the new value of Bin 2 for $S_2$ is $1 + C$.  For Bin 5, $q_5 = 1$, so we go circular right and find non-empty Bin 1 at distance 2.  The new hash value of Bin 5 is therefore $1+2C$. Note that the  non-empty bins remain unchanged.

Formally, let $q_j$ $j = \{0,1,2,...,k-1\}$ be $k$ i.i.d. Bernoulli random variables such that $q_j = 1$ with probability $\frac{1}{2}$. The improved hash function $\mathcal{H}^{+}$ is given by
\begin{align}\label{eqn_H}
&\mathcal{H}^+_j(S) =\left\{\begin{array}{lll}
\underset{(j-t_1)\text{mod}\ k}{OPH}(\pi(S)) + t_1 C &\\ & \hspace{-1.in}\text{ if } q_j = 0 \text{ and} \  \underset{j}{OPH}(\pi(S))= E\\\\
\underset{(j+t_2)\text{mod}\ k}{OPH}(\pi(S)) + t_2C &\\ &\hspace{-1.in}\text{ if } q_j = 1 \text{ and} \  \underset{j}{OPH}(\pi(S))= E \\\\
\underset{j}{OPH}(\pi(S))  &\text{otherwise}
\end{array}
\right.
\end{align}
where
\begin{align}
&t_1 = \min z,\ \ \ s.t.\ \ \underset{(j-z)\ \text{mod}\ k}{OPH}(\pi(S))\neq E\\
&t_2 = \min z,\ \ \ s.t.\ \ \underset{(j+z)\ \text{mod}\ k}{OPH}(\pi(S))\neq E
\end{align}
with same  $C = \frac{D}{k}+1$.
Computing $k$ hash evaluations with $\mathcal{H}^+$ requires evaluating $\pi(S)$ followed by two passes over the $k$ bins from different directions. The total complexity of computing $k$ hash evaluations is again $O(d + k)$ which is the same as that of the existing densification scheme. We need an additional storage of the $k$ bits (roughly hundreds or thousands in practice) which is practically negligible.

It is not difficult to show that $\mathcal{H}^+$ satisfies the LSH property for resemblance, which we state as a theorem.
\begin{theorem}\label{thm_H_Pr}
\begin{align}
\mathbf{Pr}\left(\mathcal{H}^+_j(S_1) = \mathcal{H}^+_j(S_2)\right) = R
\end{align}
\end{theorem}
$\mathcal{H}^+$ leads to an unbiased estimator of resemblance $\hat{R}^+$
\begin{equation}
\hat{R}^+ = \frac{1}{k}\sum_{j=0}^{k-1} \mathbf{1}\{\mathcal{H}^+_{j}(S_1) = \mathcal{H}^+_{j}(S_2)\}.
\end{equation}

\vspace{-0.1in}
\section{Variance Analysis of  Improved Scheme}
\vspace{-0.05in}

When $m =1$ (an event with prob $\left(\frac{1}{k}\right)^{f_1 +f_2 -a} \simeq 0$), i.e., only one simultaneously non-empty bin, both the schemes are exactly same. For simplicity of expressions, we will assume that the number of simultaneous non-empty bins is strictly greater than 1, i.e., $m > 1$. The general case has an extra term for $m =1$, which makes the expression unnecessarily complicated without changing the final conclusion.

Following the  notation as in Sec.~\ref{sec:var_old}, we denote
\begin{align}
M_j^{N+} &= \mathbf{1}\{I_{emp}^j =0 \mbox{ and } \mathcal{H}^+_j(S_1) = \mathcal{H}^+_j(S_2)\} \\
M_j^{E+} &= \mathbf{1}\{I_{emp}^j =1 \mbox{ and } \mathcal{H}^+_j(S_1) = \mathcal{H}^+_j(S_2)\}\end{align}

The two expectations $ \mathbb{E}\left[\sum_{i \ne j}M^{N+}_i M^{N+}_j\bigg|m\right]$ and $ \mathbb{E}\left[\sum_{i \ne j}M^{N+}_i M^{E+}_j\bigg|m\right]$  are the same as given by Lemma~\ref{lem:nonemptynonempty} and Lemma~\ref{lem:emptyandnonempty} respectively, as all the arguments used to prove them still hold   for the new scheme.
 The only change is in the term $\mathbb{E}\left[\sum_{i \ne j}M^E_i M^E_j\bigg|m\right]$.

\begin{lemma}
\begin{align}
 &\mathbb{E}\left[\sum_{i \ne j}M^{E+}_i M^{E+}_j\bigg|m\right]  =  (k-m)(k-m-1) \notag\\
&\hspace{0.8in}\times \left[\frac{3R}{2(m+1)} + \frac{(2m-1)R\tilde{R}}{2(m+1)}\right]
\end{align}\end{lemma}

The theoretical variance of the new estimator $\hat{R}^+$ is given by the following Theorem~\ref{theo:var_imp}.
\begin{theorem}
\label{theo:var_imp}
\begin{align}
&Var(\hat{R}^+) =\frac{R}{k} + A^+\frac{R}{k^2} + B^+\frac{R\tilde{R}}{k^2} - R^2 \\
&A^+ = \mathbb{E}\left[\frac{N_{emp} (4k - N_{emp} +1)}{2(k - N_{emp} +1)}\right]\notag\\
&B^+ =\mathbb{E}\left[\frac{2k^3  + N_{emp}^2 - N_{emp}(2k^2 + 2k +1) -2k}{2(k - N_{emp} +1)}\right]\notag
\end{align}
\end{theorem}

\begin{figure*}[ht]
\begin{center}

\mbox{
\includegraphics[width = 1.8in]{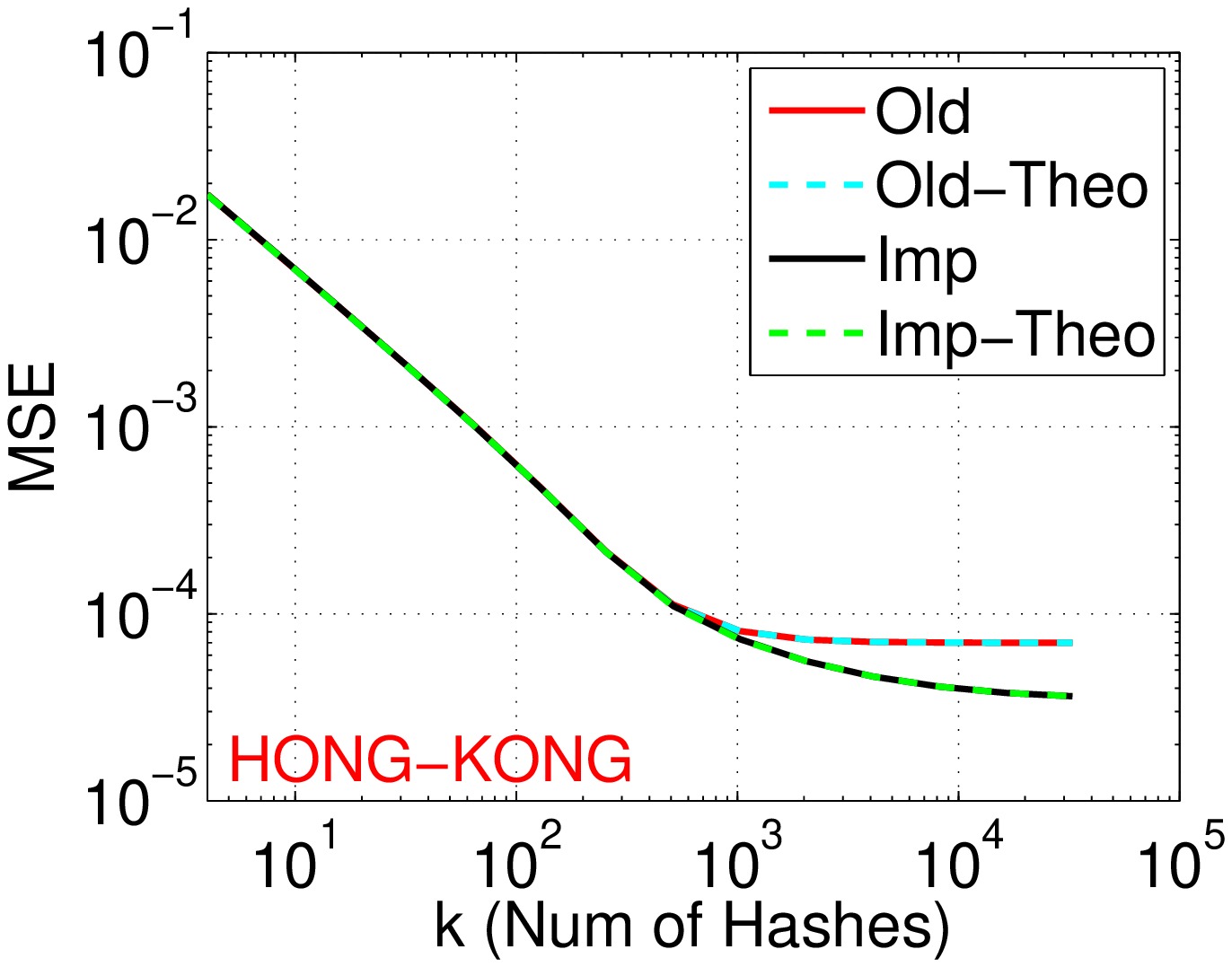}\hspace{-0.15in}
\includegraphics[width = 1.8in]{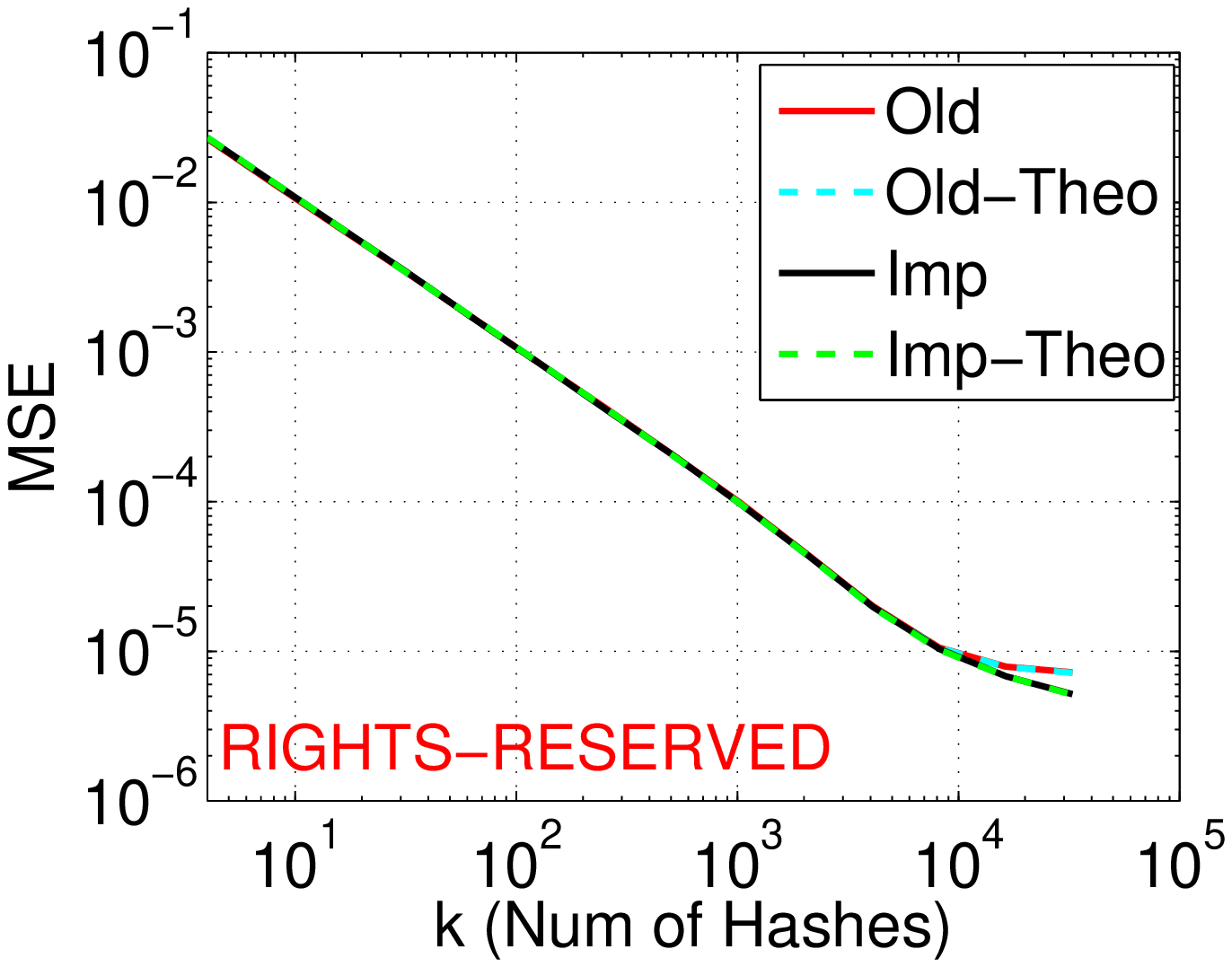}\hspace{-0.15in}
\includegraphics[width = 1.8in]{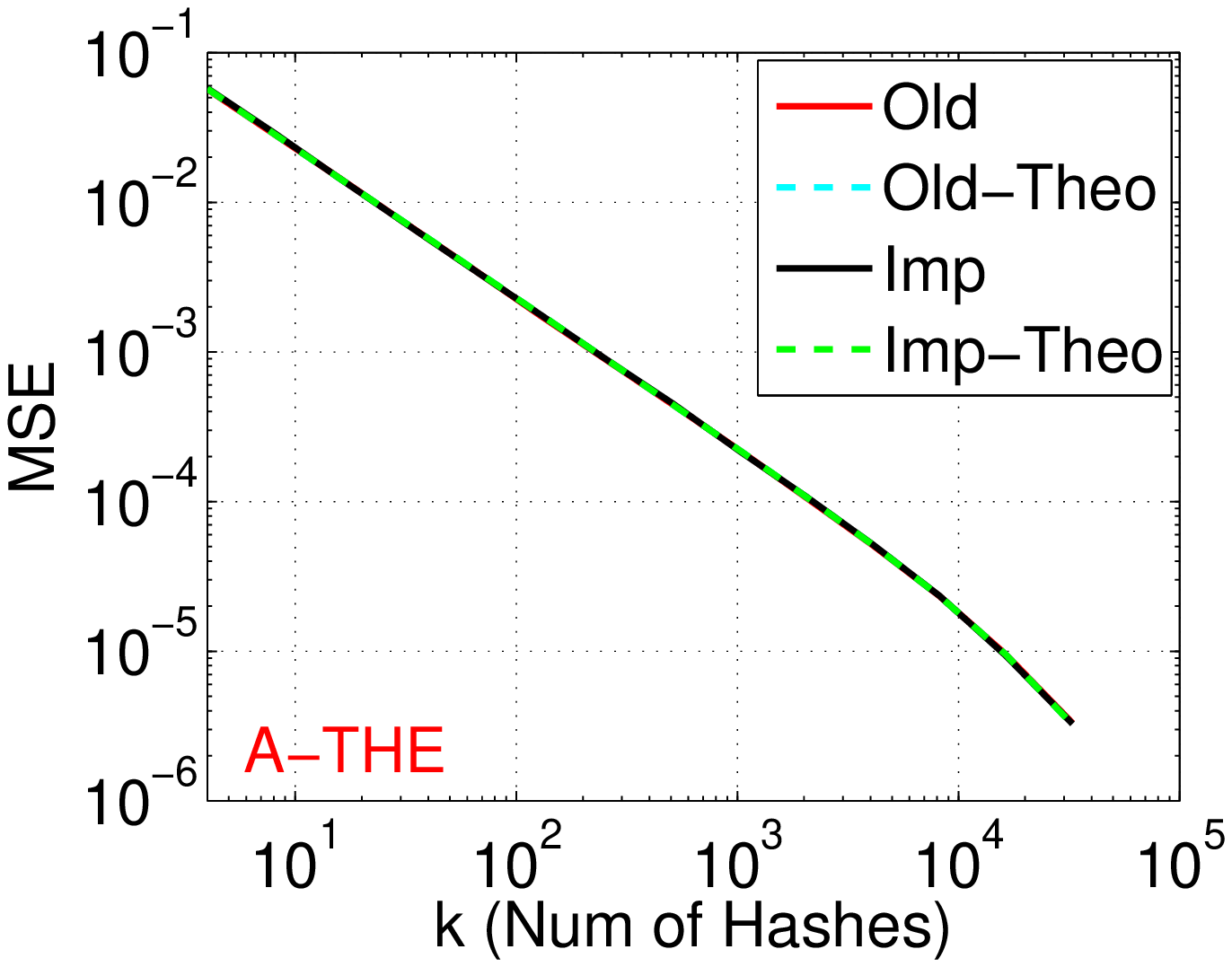}\hspace{-0.15in}
\includegraphics[width = 1.8in]{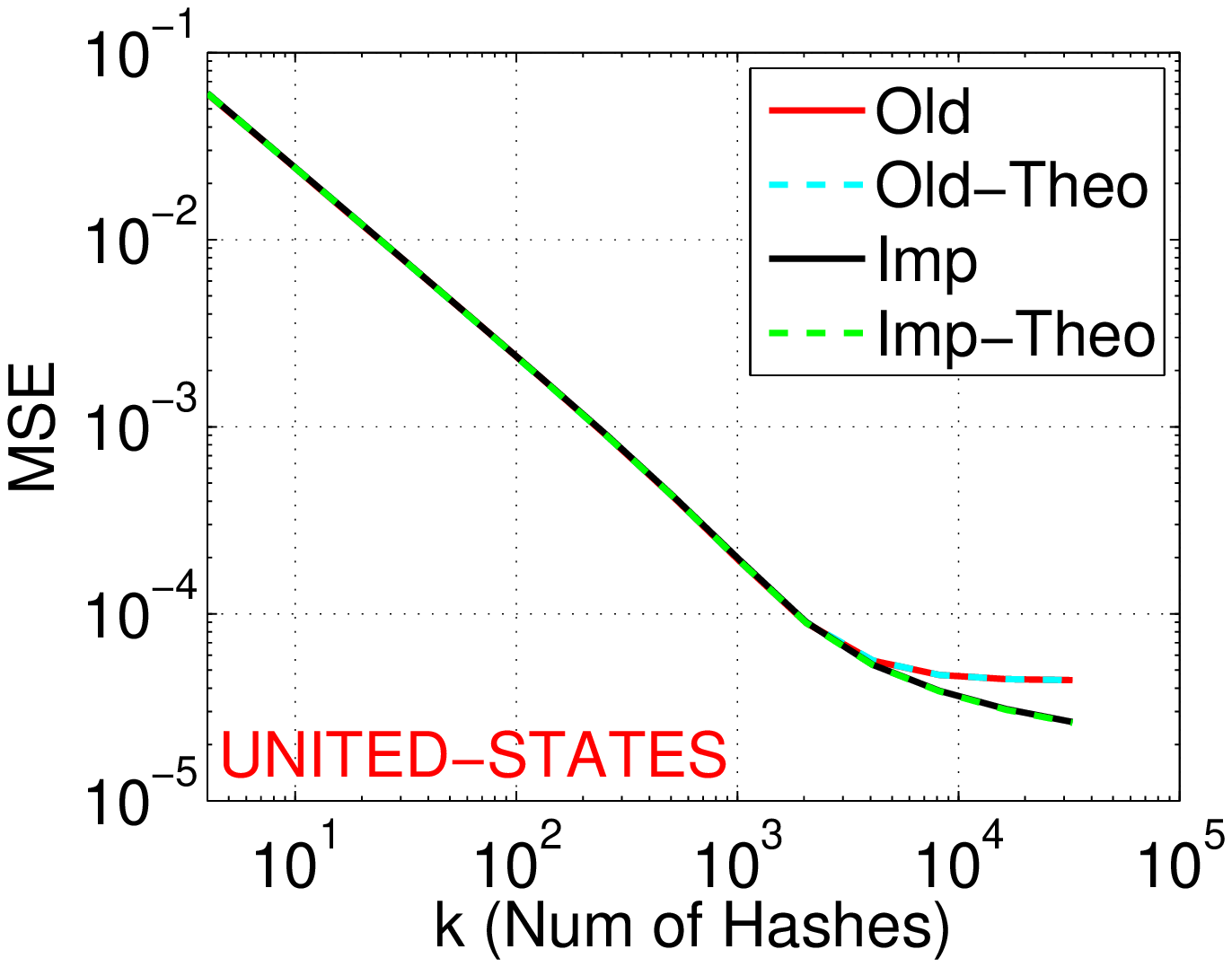}
}

\mbox{
\includegraphics[width = 1.8in]{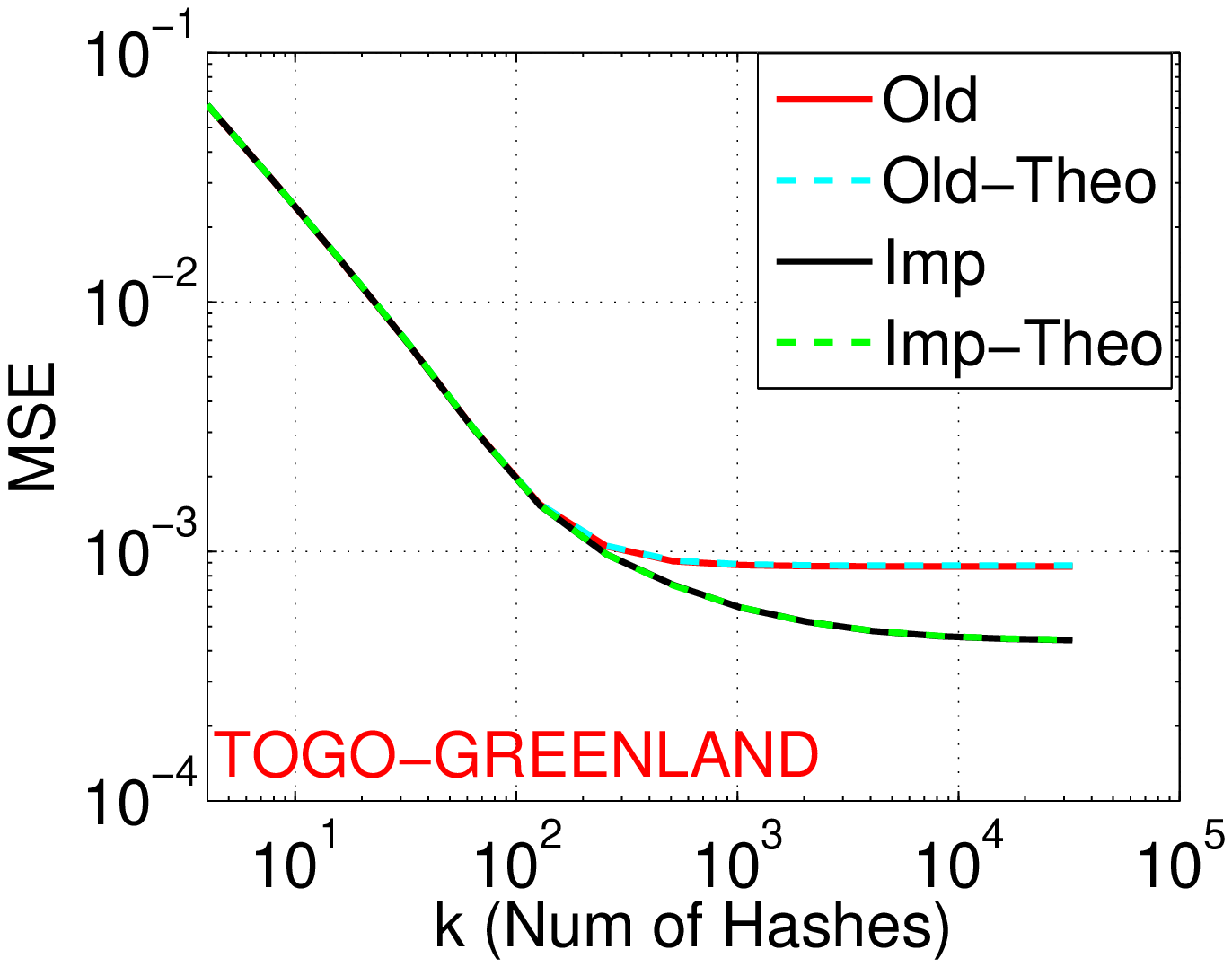}\hspace{-0.15in}
\includegraphics[width = 1.8in]{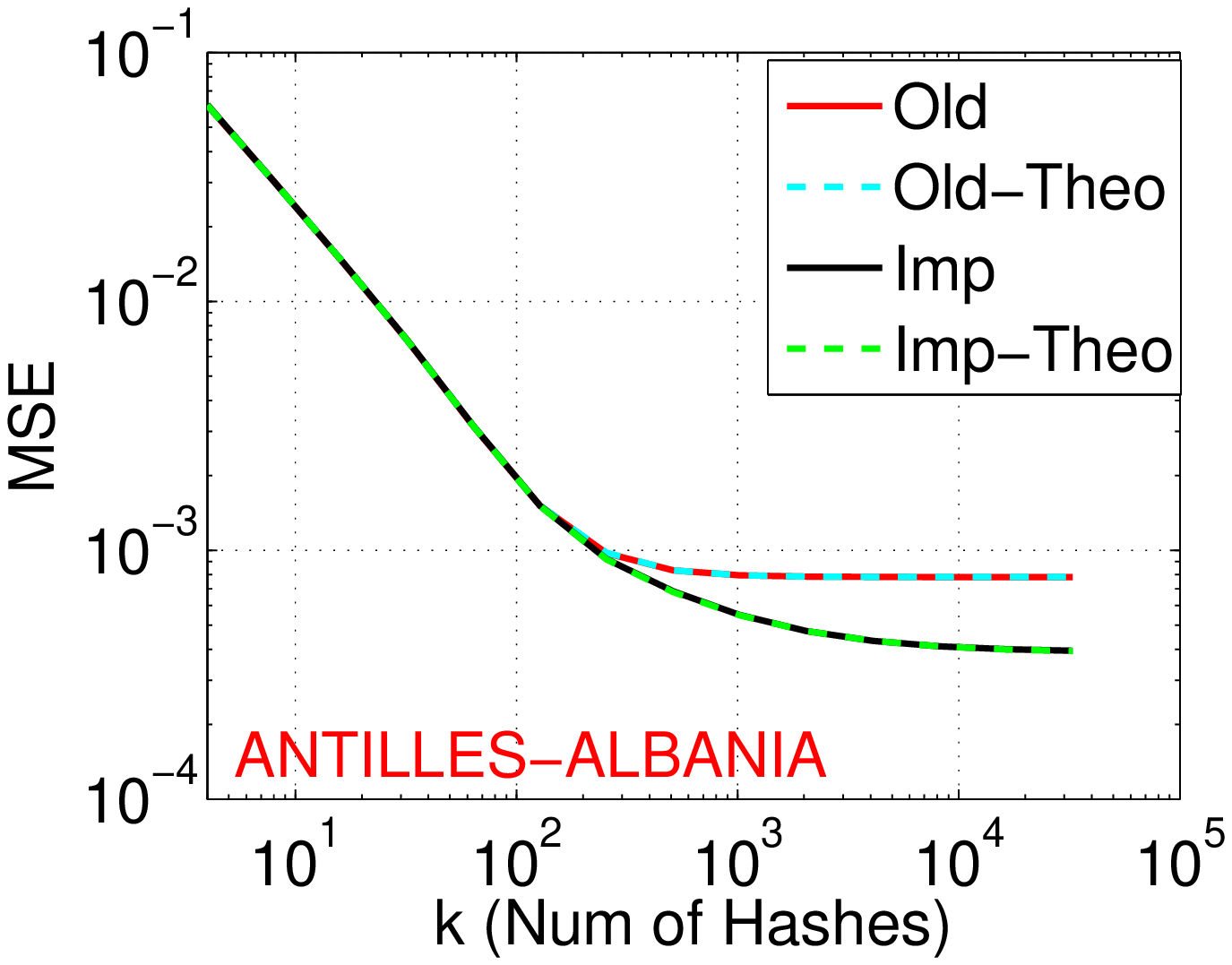}\hspace{-0.15in}
\includegraphics[width = 1.8in]{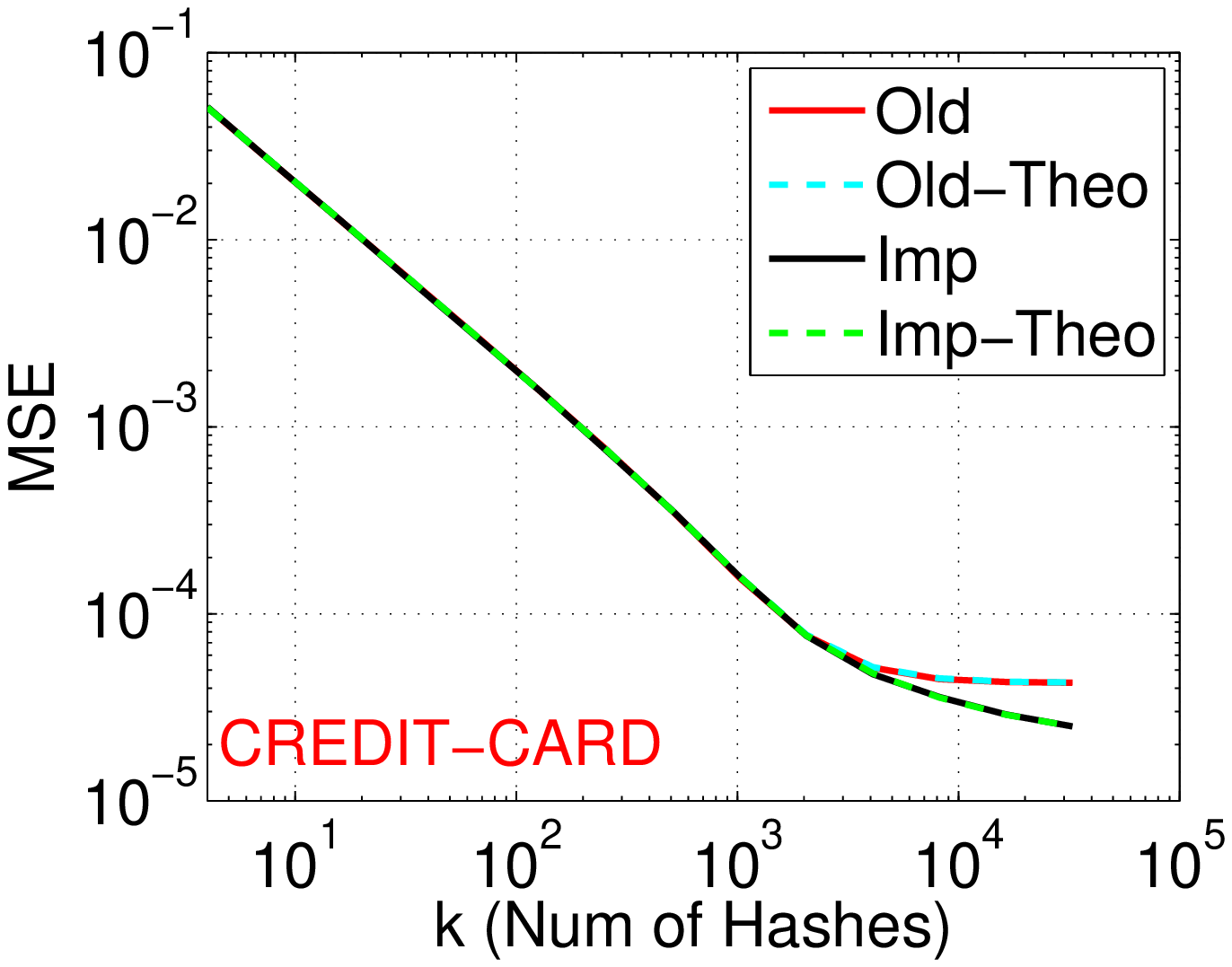}\hspace{-0.15in}
\includegraphics[width = 1.8in]{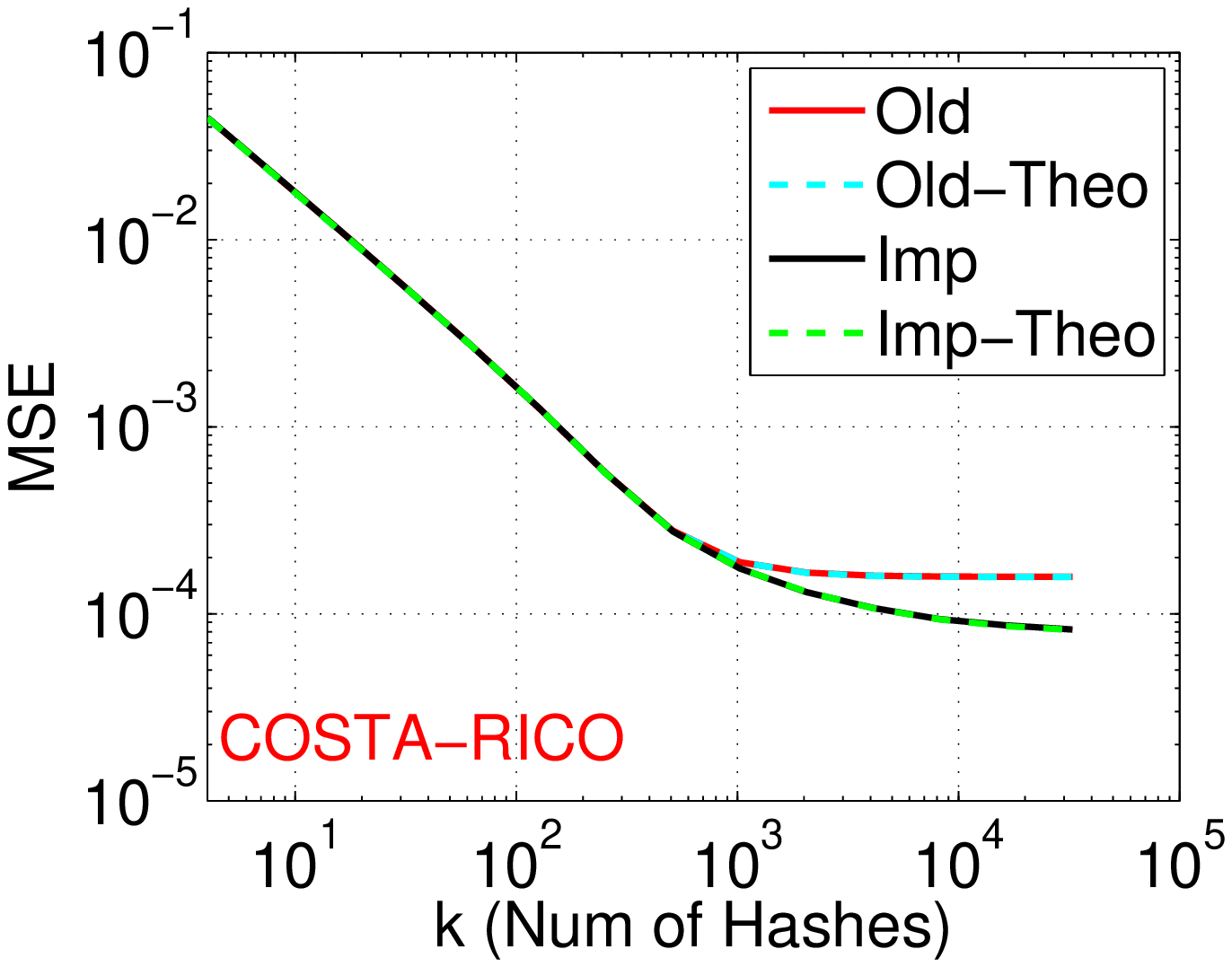}
}
\mbox{
\includegraphics[width = 1.8in]{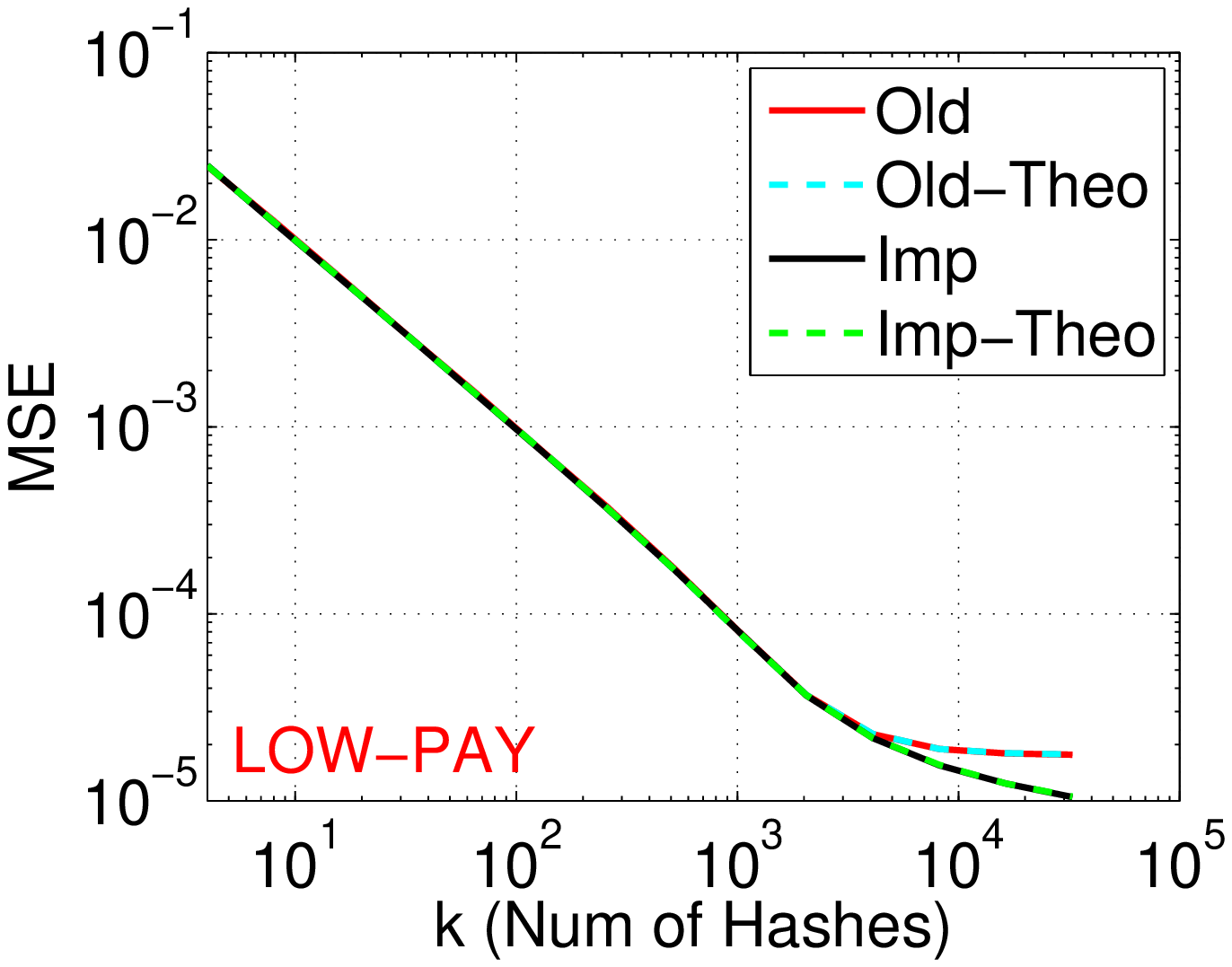}\hspace{-0.15in}
\includegraphics[width = 1.8in]{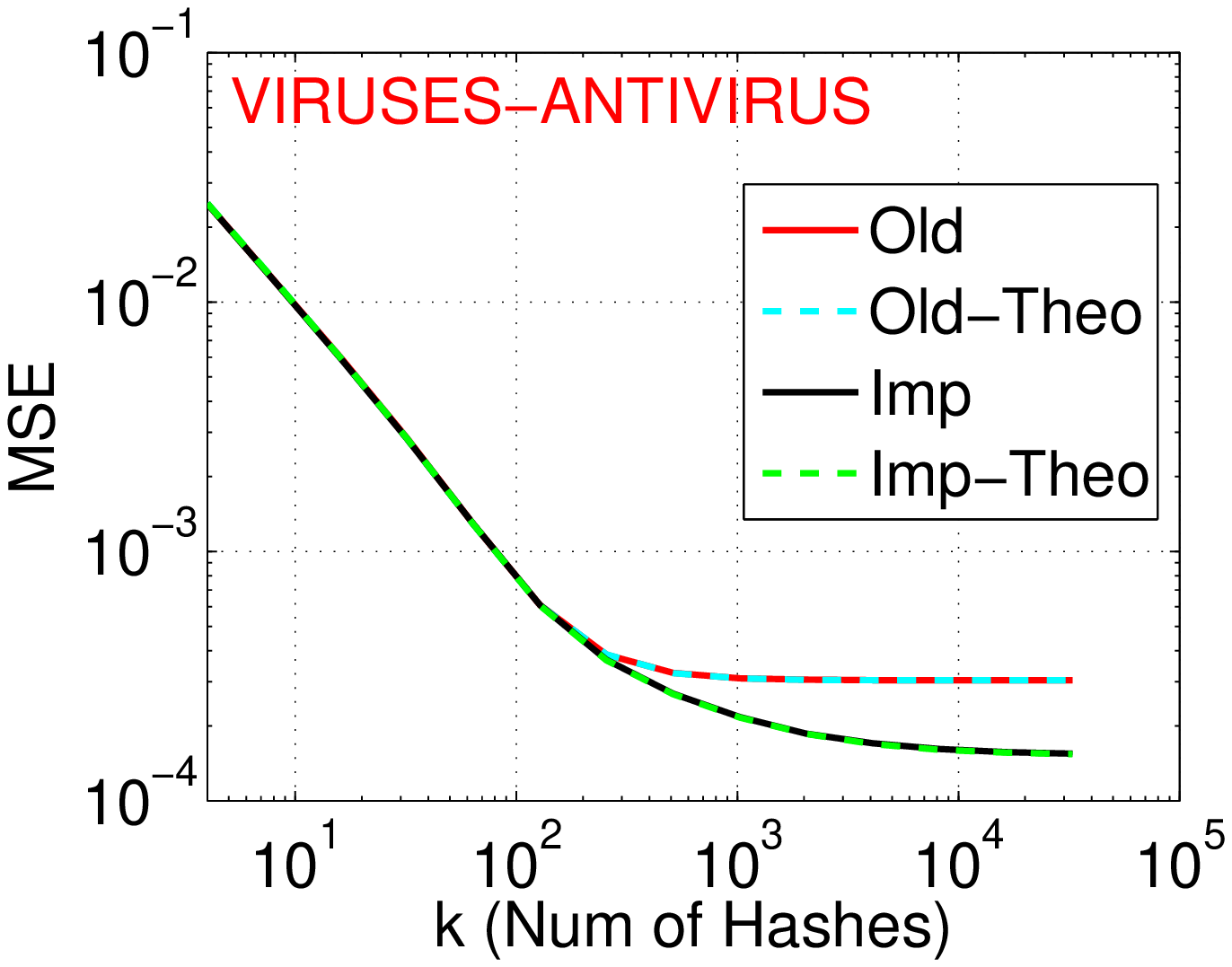}\hspace{-0.15in}
\includegraphics[width = 1.8in]{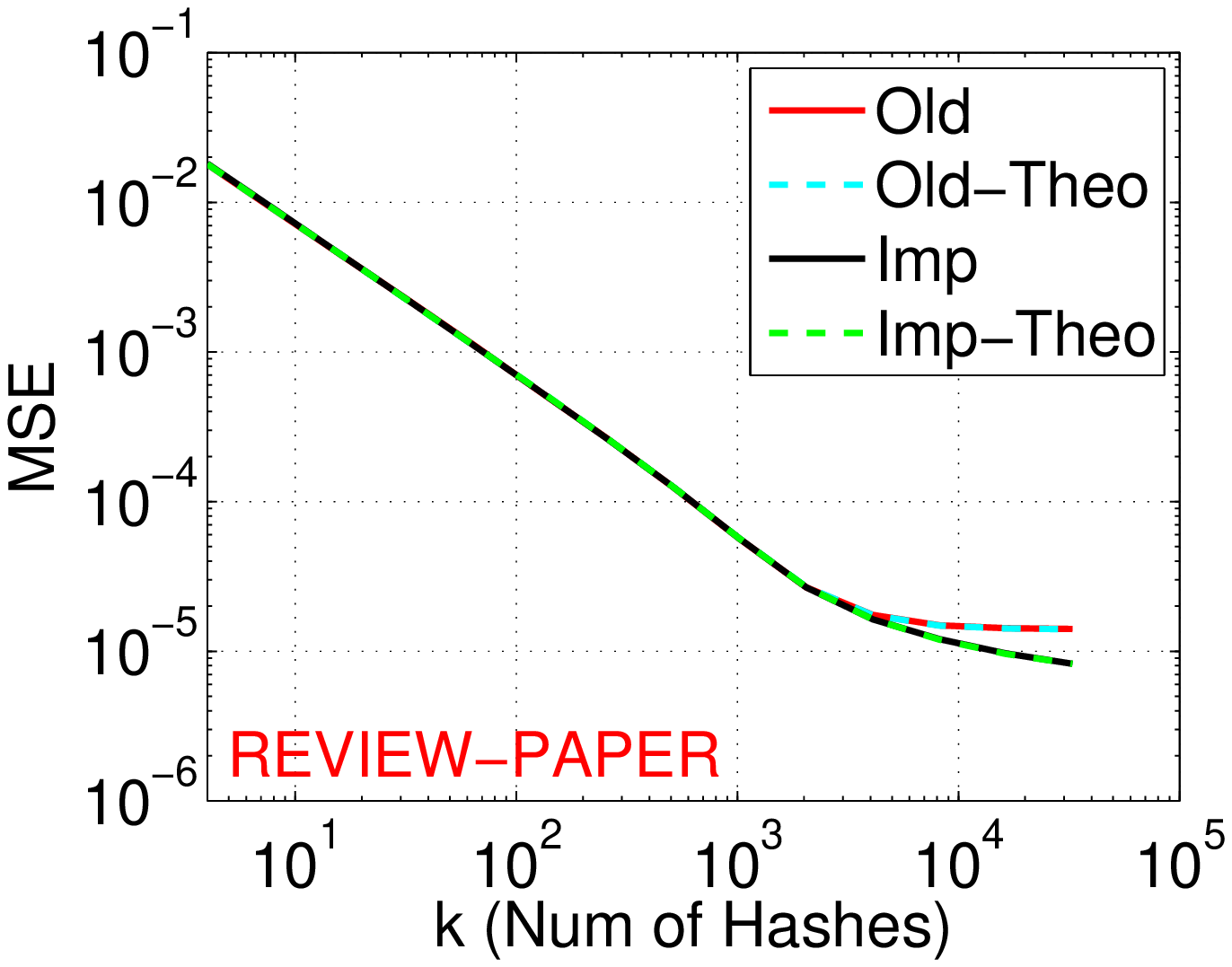}\hspace{-0.15in}
\includegraphics[width = 1.8in]{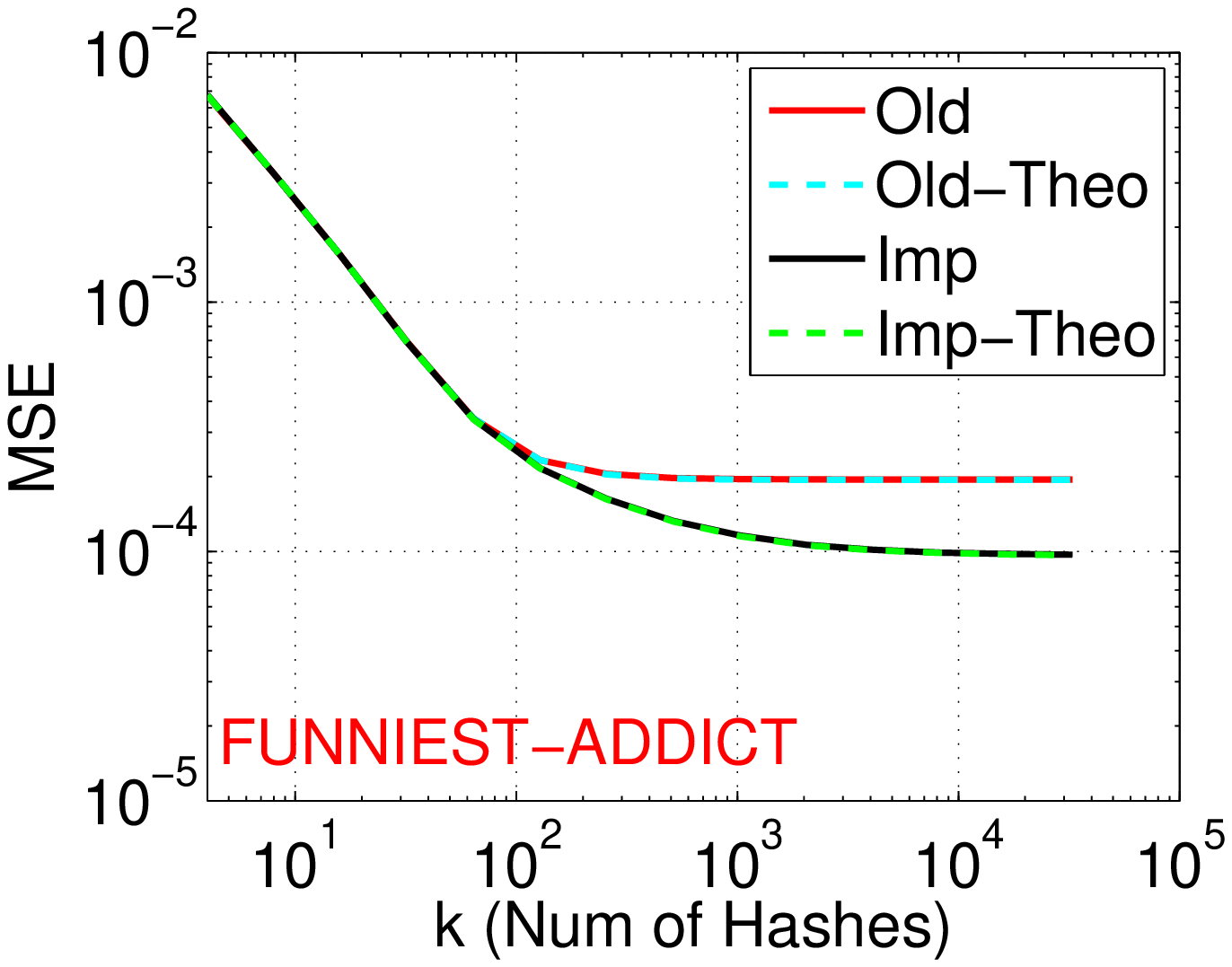}
}

\end{center}
\vspace{-0.2in}
\caption{Mean Square Error (MSE) of the old scheme $\hat{R}$ and the improved scheme $\hat{R}^+$ along with their theoretical values on 12 word pairs (Table~\ref{tab_12pairs}) from a web crawl dataset.}\label{fig:MSE}\vspace{-0.1in}
\end{figure*}

The new scheme reduces the value of $p$ (see Eq.(\ref{eq:p})) from $\frac{2}{m+1}$ to $\frac{1.5}{m+1}$. As argued in Sec.~\ref{sec:intut},  this reduces the overall variance. Here, we state it as theorem that $Var(\hat{R}^+) \le Var(\hat{R})$ always.
 \begin{theorem}
 \label{theo:main_ineq}
\begin{align}
Var(\hat{R}^+) \le Var(\hat{R})
\end{align}
More precisely,
\begin{align}\notag
&Var(\hat{R}) -  Var(\hat{R}^+)\\
 =&   \mathbb{E}\left[\frac{(N_{emp})(N_{emp}-1)}{2k^2(k -N_{emp}+1)}[R - R\tilde{R}]\right]
\end{align}
\end{theorem}

The probability of simultaneously empty bins increases with increasing sparsity in dataset and the total number of bins $k$. We can see from Theorem~\ref{theo:main_ineq} that with more simultaneously empty bins, i.e., higher  $N_{emp}$, the gain with the improved scheme $\mathcal{H}^{+}$ is higher compared to $\mathcal{H}$. Hence, $\mathcal{H}^+$ should be significantly better than the existing scheme for very sparse datasets or in scenarios when we need a large number of hash values.

%
\vspace{-0.1in}
\section{Evaluations}
\vspace{-0.05in}

Our first experiment concerns the validation of the theoretical variances of the two densification schemes. The second experiment focuses on comparing the two schemes in the context of near neighbor search with LSH.

\vspace{-0.1in}
\subsection{Comparisons of Mean Square Errors}
\vspace{-0.05in}

We empirically verify the theoretical variances of $\mathcal{R}$ and $\mathcal{R}^+$ and their effects in many practical scenarios.  To achieve this, we extracted 12 pairs of words  (which cover a wide spectrum of sparsity and similarity) from the web-crawl dataset which consists of word representation from  $2^{16}$ documents.  Every word is represented as a binary vector (or set) of $D = 2^{16}$ dimension, with a feature value of 1 indicating the presence of that word in the corresponding document.  See Table~\ref{tab_12pairs} for  detailed information of the data.

\begin{table}[h]
\caption{Information of 12 pairs of word vectors. Each word stands for  a set of documents in which the word is contained. For example, ``A'' corresponds to the set of document IDs which contained  word ``A''.
 }
\begin{center}{\small
\begin{tabular}{l l r r r  }
\hline \hline
Word 1 & Word 2 &$f_1$  &$f_2$  &$R$  \\\hline
HONG & KONG &940   &948 &0.925\\
RIGHTS & RESERVED &12,234       &11,272 &0.877\\
A & THE &39,063     &42,754 &0.644\\
UNITED &STATES & 4,079        &3,981 &0.591\\
TOGO &GREENLAND &231 &200 &0.528\\
ANTILLES &ALBANIA &184 &275  &0.457\\
CREDIT &CARD &2,999        &2,697 &0.285\\
COSTA & RICO &773 &611 &0.234 \\
LOW &PAY &2,936        &2,828 &0.112\\%
VIRUSES &ANTIVIRUS  &212 &152 &0.113 \\
REVIEW & PAPER &3,197        &1,944  &0.078\\
FUNNIEST &ADDICT &68 &77 &0.028\\
\hline\hline
\end{tabular}
}
\end{center}
\label{tab_12pairs}\vspace{-0.15in}
\end{table}

For all  12 pairs of words, we estimate the resemblance using the two estimators $\mathcal{R}$ and $\mathcal{R}^+$. We plot the empirical {\em Mean Square Error (MSE)}  of both estimators with respect to $k$ which is the number of hash evaluations. To validate the theoretical variances (which is also the MSE because the estimators are unbiased), we also plot the values of the theoretical variances computed from Theorem~\ref{theo:var_old} and Theorem~\ref{theo:var_imp}.  The results are summarized in Figure~\ref{fig:MSE}.

From the plots we can see that the theoretical and the empirical MSE values overlap in both the cases validating both Theorem~\ref{theo:var_old} and Theorem~\ref{theo:var_imp}. When $k$ is small both the schemes have similar variances, but when $k$ increases the improved scheme always shows better variance. For very sparse pairs, we start seeing a significant difference in variance even for $k$ as small as 100. For  a sparse pair, e.g., ``TOGO" and ``GREENLAND", the difference in variance, between the two schemes,  is more compared to the dense pair  ``A" and ``THE". This is in agreement with Theorem~\ref{theo:main_ineq}.

\vspace{-0.1in}
\subsection{Near Neighbor Retrieval with LSH}
\vspace{-0.05in}

In this experiment, we evaluate the two hashing schemes $\mathcal{H}$ and $\mathcal{H}^+$ on the standard $(K,L)$-parameterized LSH algorithm~\cite{Proc:Indyk_STOC98,Report:E2LSH} for retrieving near neighbors. Two publicly available sparse text datasets are described in Table~\ref{tab_data}.

\begin{table}[h]
\caption{ Dataset information. \vspace{-0.2in}}
\begin{center}{
\begin{tabular}{l r r r r}
\hline \hline
Data &\# dim &\# nonzeros &\# train &\# query\\\hline
RCV1 &47,236 &73 &100,000 &5,000\\
URL & 3,231,961 &115 &90,000 &5,000 \\
\hline\hline
\end{tabular}
}
\end{center}
\label{tab_data}\vspace{-0.1in}
\end{table}

In $(K,L)$-parameterized LSH algorithm for near neighbor search, we generate $L$ different meta-hash functions. Each of these meta-hash functions is formed by concatenating $K$ different hash values as
\begin{equation}
\label{eq:bucket}
B_j(S)  = [h_{j1}(S);h_{j2}(S);...;h_{{jK}}(S)],
\end{equation}
 where $h_{ij}, i \in \{1,2,...,K \}$, $j \in  \{1,2,...,L \}$, are $KL$  realizations of the hash function under consideration. The $(K,L)$-parameterized LSH works in two phases:
\begin{enumerate}
\item {\bf Preprocessing Phase:} We construct $L$ hash tables from the data by storing element $S$, in the train set, at location $B_j(S)$ in hash-table $j$.
\item {\bf Query Phase:} Given a query $Q$, we report the union of all the points in the buckets $B_j(Q)$ $\forall j \in \{1,2,...,L\}$, where the union is over $L$ hash tables.
\end{enumerate}

For every dataset, based on the similarity levels, we chose  a  $K$ based on standard recommendation. For this  $K$ we show results for a set of values of $L$ depending on the recall values. Please refer to~\cite{Report:E2LSH} for  details on the implementation of LSH. Since both $\mathcal{H}$ and $\mathcal{H}^+$ have the same collision probability, the choice of $K$ and $L$ is the same in both  cases.

\begin{figure}[ht]
\begin{center}

\mbox{
\includegraphics[width = 1.7in]{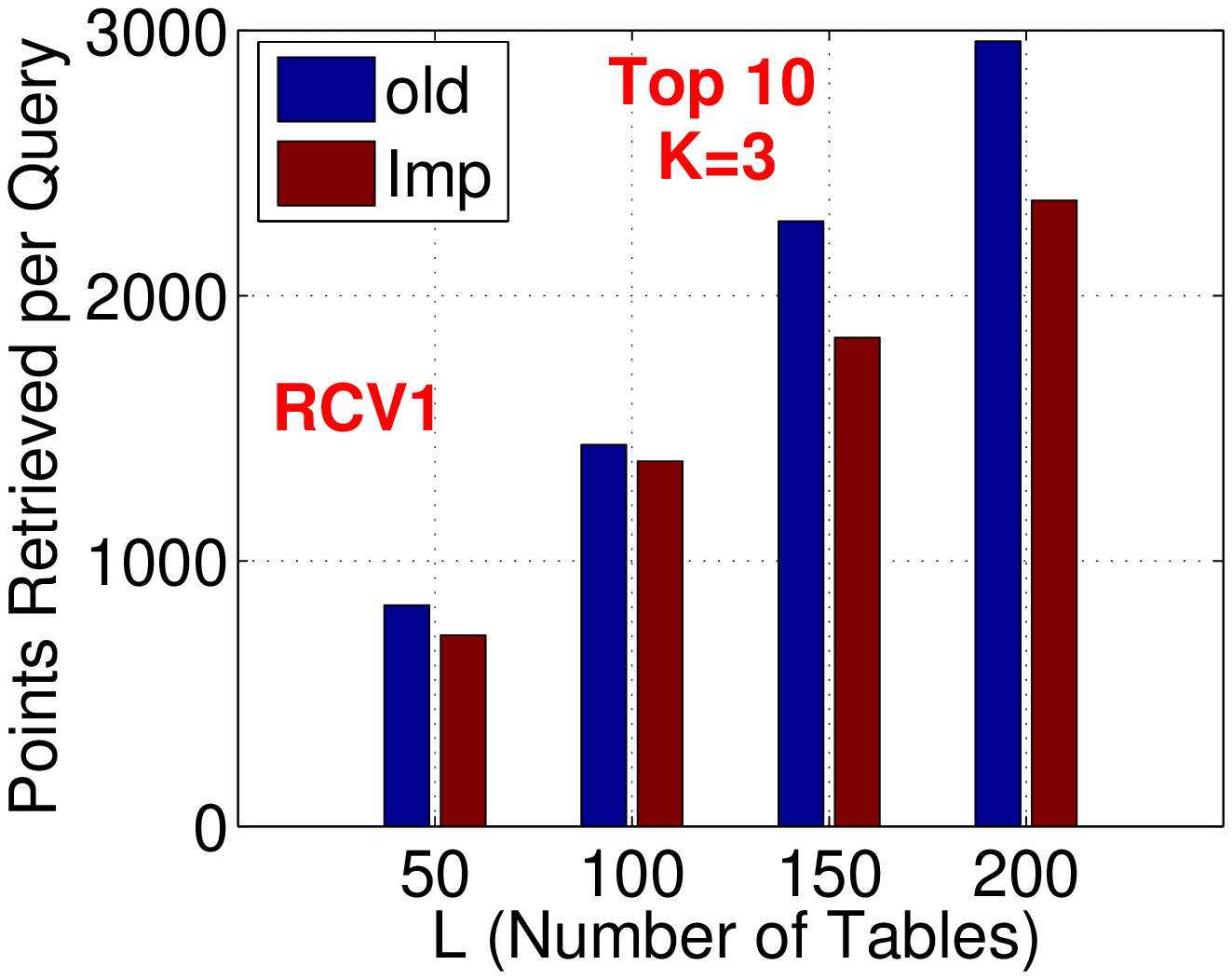}\hspace{-0.1in}
\includegraphics[width = 1.7in]{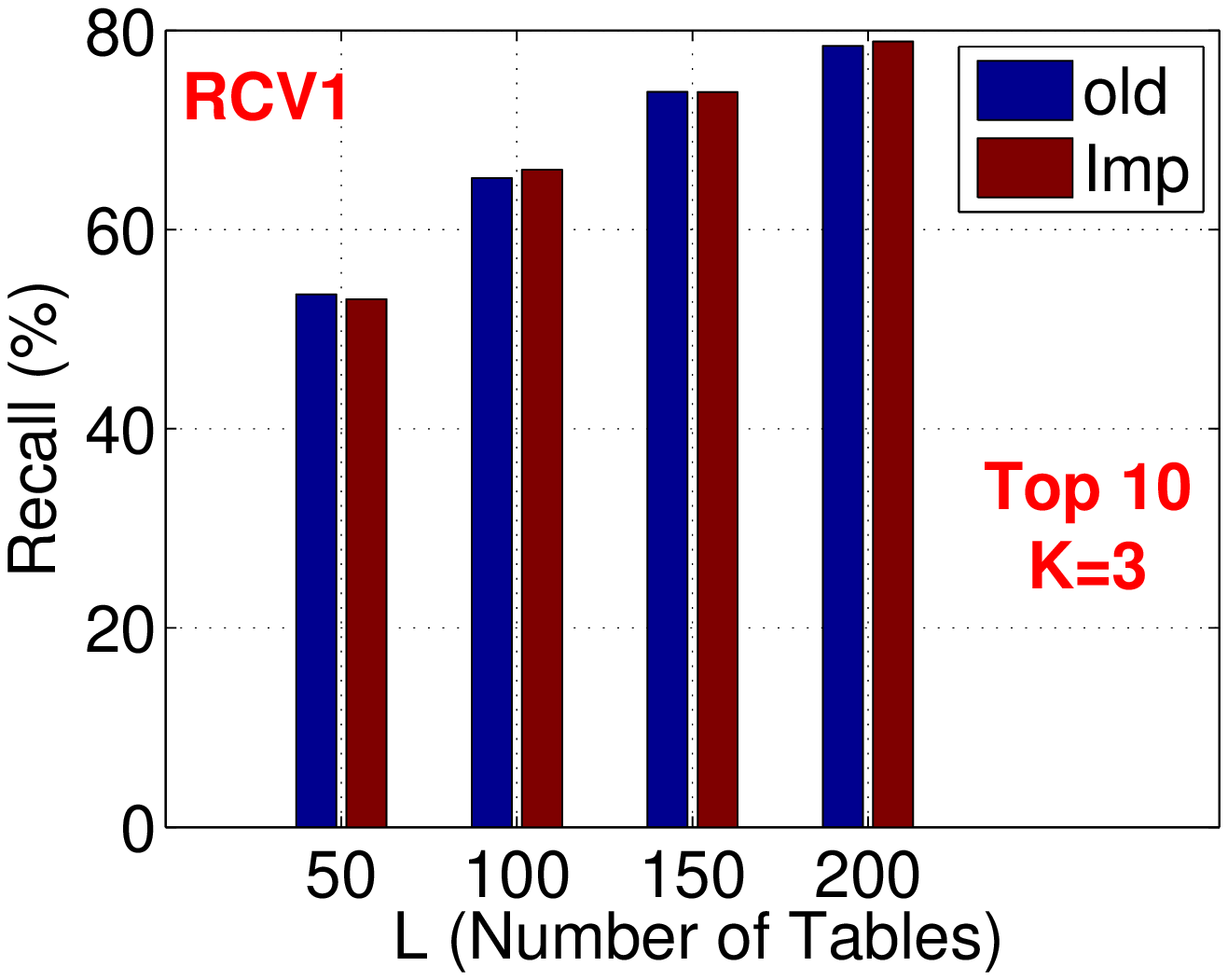}
}

\mbox{
\includegraphics[width = 1.7in]{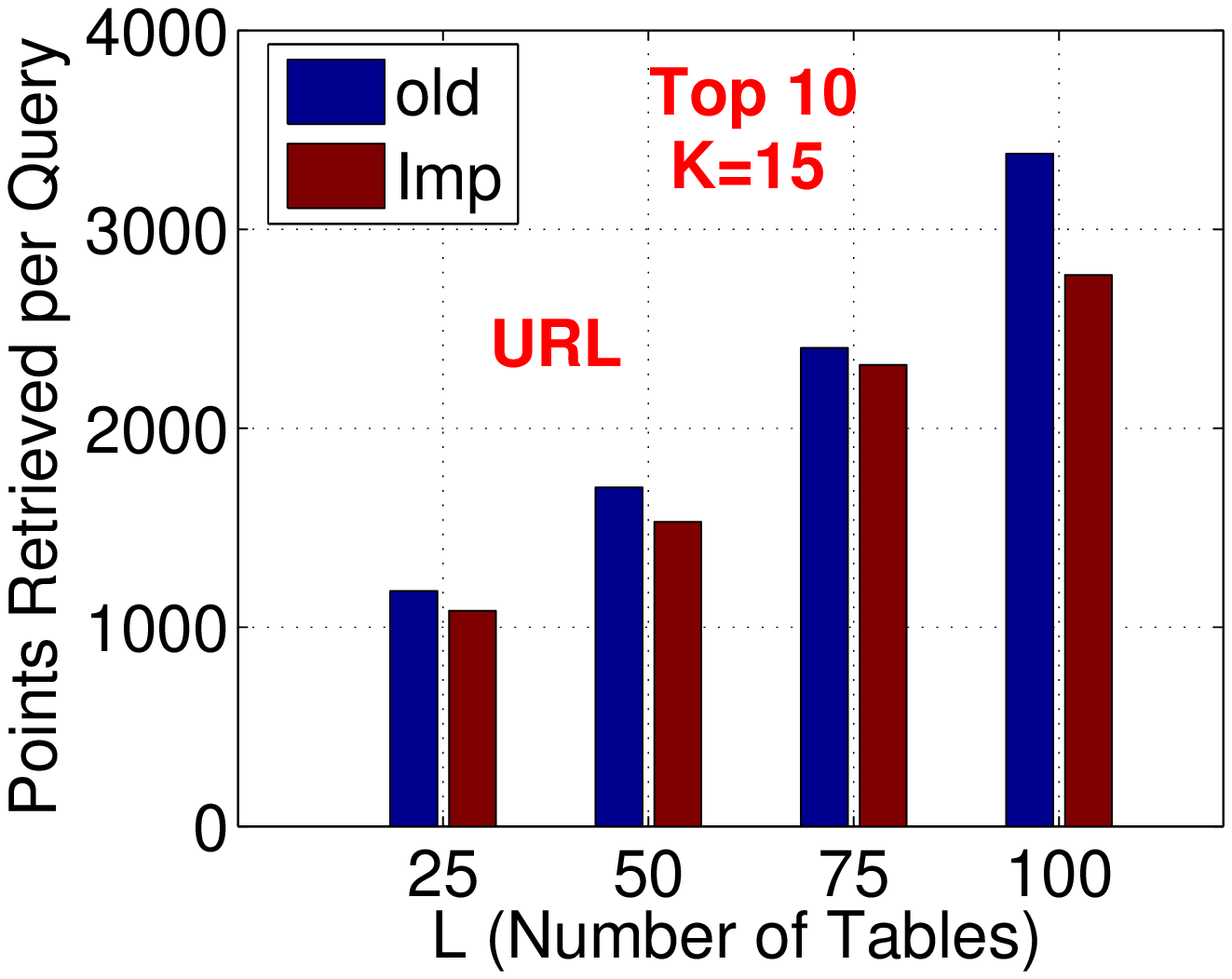}\hspace{-0.1in}
\includegraphics[width = 1.7in]{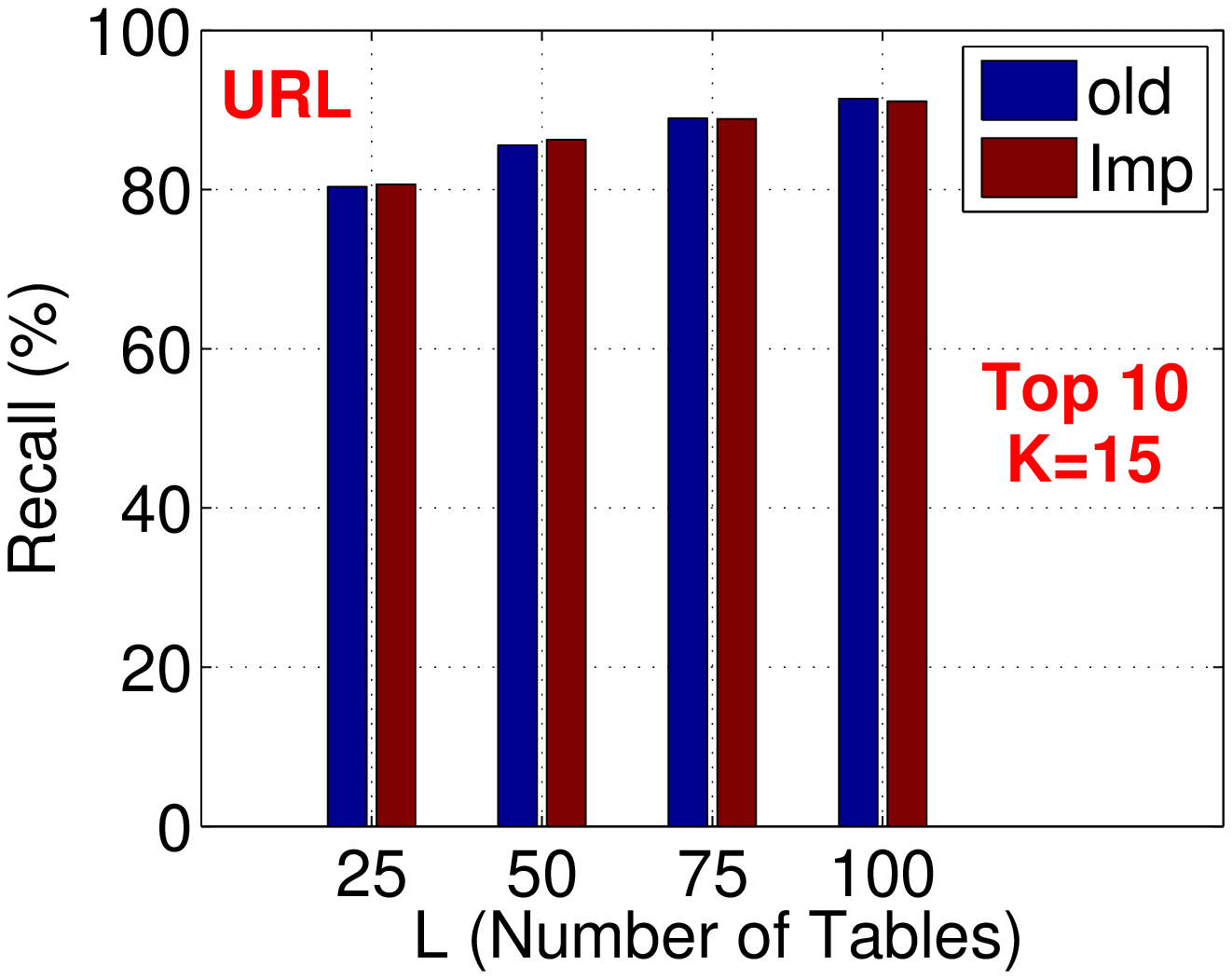}

}

\end{center}
\vspace{-0.1in}
\caption{ Average number of points scanned per query and the mean recall values of top 10 near neighbors, obtained from $(K,L)$-parameterized LSH algorithm, using   $\mathcal{H}$ (old) and $\mathcal{H}^+$ (Imp). Both schemes achieve the same recall but $\mathcal{H}^+$  reports fewer points compared to $\mathcal{H}$. Results are averaged over 10 independent runs.    }\label{fig:NN_expts}
\end{figure}

For every query point, the gold standard  top 10 near neighbors from the training set are computed based on actual resemblance. We then compute the recall  of these gold standard neighbors and the total number of points retrieved by the $(K,L)$ bucketing scheme. We report the mean computed over all the points in the query set. Since the experiments involve randomization, the final results presented are averaged over 10 independent runs. The recalls and the points retrieved per query are summarized in  Figure~\ref{fig:NN_expts}.

It is clear from Figure~\ref{fig:NN_expts} that the improved hashing scheme $\mathcal{H}^+$ achieves the same recall but at the same time retrieves  less number of points compared to the old scheme $\mathcal{H}$. To achieve $90\%$ recall on URL dataset, the old scheme retrieves around 3300 points per query on an average while the improved scheme only needs to check around 2700 points per query. For RCV1 dataset, with $L=200$ the old scheme retrieves around 3000 points and achieves a recall of $80\%$, while the same recall is achieved by the improved scheme after retrieving only about 2350 points per query.  A good hash function provides a right balance between recall and number of points retrieved. In particular, a hash function which achieves a given recall and at the same time retrieves less number of points is desirable because it implies better precision.  The above results clearly demonstrate the superiority of the indexing scheme with improved hash function  $\mathcal{H}^+$ over the indexing scheme with $\mathcal{H}$.

\vspace{-0.1in}
\subsection{Why $\mathcal{H}^+$  retrieves less number of points than  $\mathcal{H}$ ?}\label{sec:kmoment_why}
\vspace{-0.05in}

The number of points retrieved, by the $(K,L)$ parameterized LSH algorithm,  is directly related to the collision probability of the meta-hash function $B_j(.)$ (Eq.(\ref{eq:bucket})). Given $S_1$ and $S_2$ with resemblance $R$, the higher the probability of event $B_j(S_1) = B_j(S_2)$, under a hashing scheme, the more number of points  will be retrieved per table.

The analysis of the variance (second moment) about the event $B_j(S_1) = B_j(S_2)$ under $\mathcal{H}^+$ and $\mathcal{H}$ provides some reasonable insight. Recall that since both estimators under the two hashing schemes are unbiased, the analysis of the first moment does not provide information in this regard.
\begin{align}
&\mathbb{E}\big[{1}\{\mathcal{H}_{j1}(S_1) =\mathcal{H}_{j1}(S_2)\} \times {1}\{\mathcal{H}_{j2}(S_1) =\mathcal{H}_{j2}(S_2)\}\big]\notag\\
&= \mathbb{E}\left[M_{j1}^N M_{j2}^N + M_{j1}^N M_{j2}^E + M_{j1}^E M_{j2}^N + M^E_{j1} M^E_{j2}\right]\notag
\end{align}

As we know from our analysis that the first three terms inside expectation, in the RHS of the above equation, behaves similarly for both $\mathcal{H}^+$ and $\mathcal{H}$. The fourth term $\mathbb{E}\left[M^E_{j1} M^E_{j2} \right]$ is likely to be smaller in case of $\mathcal{H^+}$ because of smaller values of $p$. We therefore see that $\mathcal{H}$  retrieves more points than necessary as compared to $\mathcal{H^+}$. The difference is  visible when empty bins dominate and $M^E_1 M^E_2 = 1$ is more likely. This happens in the case of sparse datasets which are common in practice.

\vspace{-0.1in}
\section{Conclusion}
\vspace{-0.05in}

Analysis of the densification scheme for one permutation hashing, which reduces the processing time of minwise hashes, reveals a sub-optimality in the existing procedure. We provide a simple improved procedure which adds more randomness in the current densification technique leading to a provably better scheme, especially for very sparse datasets. The improvement comes without any compromise with the computation and only requires $O(d + k)$ (linear) cost for generating $k$ hash evaluations.
We hope that our improved scheme will be adopted in practice.

\vspace{-0.1in}
\section*{Acknowledgement}
\vspace{-0.05in}

Anshumali Shrivastava is a Ph.D. student  partially supported by  NSF (DMS0808864, III1249316) and ONR (N00014-13-1-0764). The work of Ping Li is partially  supported by AFOSR (FA9550-13-1-0137), ONR (N00014-13-1-0764), and NSF (III1360971, BIGDATA1419210).

\appendix

\vspace{-0.1in}
\section{Proofs}
\vspace{-0.05in}

For the analysis, it is sufficient to consider the configurations, of empty and non-empty bins, arising after throwing $|S_1 \cup S_2|$ balls uniformly into $k$ bins with exactly $m$ non-empty bins and $k - m$ empty bins. Under uniform throwing of balls, any ordering of $m$ non-empty and $k-m$ empty bins is equally likely. The proofs involve elementary combinatorial arguments of counting configurations.

\vspace{-0.1in}
\subsection{Proof of Lemma 1}
\vspace{-0.05in}

 Given exactly $m$ simultaneously non-empty bins, any two of them can be chosen in $m(m-1)$ ways (with ordering of $i$ and $j$). Each term $M_i^NM_j^N$, for both simultaneously non-empty $i$ and $j$, is 1 with probability $R\tilde{R}$ (Note, $ \mathbb{E}\left(M_i^NM_j^N \big| i \ne j, I_{emp}^i = 0, I_{emp}^j = 0\right) = R\tilde{R}$).

\vspace{-0.1in}
\subsection{Proof of Lemma 2}
\vspace{-0.05in}

The permutation is random and any sequence of simultaneously $m$ non-empty and remaining $k-m$ empty bins are equal likely. This is because, while randomly throwing $|S_1 \cup S_2|$ balls into $k$ bins with exactly $m$ non-empty bins every sequence of simultaneously empty and non-empty bins has equal probability. Given $m$, there are total $2m(k-m)$ different pairs of empty and non-empty bins (including the ordering). Now, for every simultaneously empty bin j, i.e., $I_{emp}^j = 1$, $M_j^E$ replicates $M_t^N$ corresponding to nearest non-empty Bin $t$ which is towards the circular right.  There are two cases we need to consider:

{\bf Case 1:} $t = i$, which has probability $\frac{1}{m}$ and $$\mathbb{E}(M_i^N M_j^E| I_{emp}^i =0,  I_{emp}^j = 1 ) = \mathbb{E}(M_i^N| I_{emp}^i =0)  = R$$
{\bf Case 2:} $t \ne i$, which has probability $\frac{m-1}{m}$ and
 \begin{align}\notag
 &\mathbb{E}(M_i^N M_j^E| I_{emp}^i =0,  I_{emp}^j = 1)\\\notag
=&\mathbb{E}(M_i^N M_t^N|t \ne i, I_{emp}^i =0, \ I_{emp}^t = 0)  = R\tilde{R}
\end{align}
Thus, the value of $\mathbb{E}\left[\sum_{i \ne j}M_i^N M_j^E\bigg|m\right]$ comes out to be
\begin{align}
 2m(k-m)\left[ \frac{ R}{m} + \frac{(m-1)R\tilde{R}}{m}\right]\notag
\end{align}
which is the desired expression.

\subsection{Proof of Lemma 3}

Given $m$, we have $(k-m)(k-m-1)$ different pairs of simultaneous non-empty bins.  There are two cases, if the closest simultaneous non-empty bins towards their circular right are identical, then for such $i$ and $j$, $M^E_i M^E_j = 1$ with probability $R$, else $M^E_i M^E_j = 1$ with probability $R\tilde{R}$. Let $p$ be the probability that two simultaneously empty bins $i$ and $j$ have the same closest bin on the right. Then $\mathbb{E}\left[\sum_{i \ne j}M^E_i M^E_j\bigg|m\right]$ is given by
\begin{align}
\label{eq}
(k-m)(k-m-1)\left[pR + (1- p)R\tilde{R}\right]
\end{align}
because with probability $(1-p)$, it uses estimators from different simultaneous non-empty bins and in that case the $M^E_i M^E_j = 1$ with probability $R\tilde{R}$.

Consider Figure~\ref{fig:cartoon_old}, where we have 3 simultaneous non-empty bins, i.e., $m =3$ (shown by colored boxes). Given any two simultaneous empty bins  Bin $i$ and Bin $j$ (out of total $k-m$)  they will occupy any of the $m+1 =4$ blank positions.  The arrow shows the chosen non-empty bins for filling the empty bins. There are $(m+1)^2 + (m+1) = (m+1)(m+2)$  different ways of fitting two simultaneous non-empty bins $i$ and $j$ between $m$ non-empty bins. Note,  if both $i$ and $j$ go to the same blank position they can be permuted. This adds  extra term $(m+1)$.

If both $i$ and $j$ choose the same blank space or the first and the last blank space, then both the simultaneous empty bins, Bin $i$ and Bin $j$, corresponds to the same  non-empty bin. The number of ways in which this happens is $2(m+1) +2 = 2(m+2)$. So, we have $$p =  \frac{2(m+2)}{(m+1)(m+2)}  = \frac{2}{m+1}.$$ Substituting $p$ in Eq.(\ref{eq}) leads to the desired expression.
\vspace{-0.1in}
\subsection{Proof of Lemma 4}
\vspace{-0.05in}

Similar to the proof of Lemma 3, we need  to compute $p$ which is the probability that two simultaneously empty bins, Bin $i$ and Bin $j$, use  information from the same bin. As argued before, the total number of positions for any two simultaneously empty bins $i$ and $j$, given $m$ simultaneously non-empty bins is $(m+1)(m+2)$. Consider Figure~\ref{fig:cartoon_new}, under the improved scheme, if both Bin $i$ and Bin $j$ choose the same blank position then they choose the same simultaneously non-empty bin with probability $\frac{1}{2}$. If Bin $i$ and Bin $j$ choose consecutive positions (e.g., position 2 and position 3) then they choose the same simultaneously non-empty bin (Bin b) with probability $\frac{1}{4}$. There are several boundary cases to consider too. Accumulating the terms leads to $$p = \frac{\frac{2(m+2)}{2} + \frac{2 m + 4}{4}}{(m+1)(m+2)} = \frac{1.5}{m+1}.$$ Substituting $p$ in Eq.(\ref{eq}) yields the desired result.

Note that $m=1$ (an event with almost zero probability) leads to the value of $p =1$. We ignore this case because it unnecessarily complicates the final expressions. $m=1$ can be easily handled and does not affect the final conclusion.

%



\newpage

\begin{thebibliography}{10}

\bibitem{Report:TeraLarning11}
A.~Agarwal, O.~Chapelle, M.~Dudik, and J.~Langford.
\newblock A reliable effective terascale linear learning system.
\newblock Technical report, arXiv:1110.4198, 2011.

\bibitem{Report:E2LSH}
A.~Andoni and P.~Indyk.
\newblock E2lsh: Exact euclidean locality sensitive hashing.
\newblock Technical report, 2004.

\bibitem{Proc:Bayardo_WWW07}
R.~J. Bayardo, Y.~Ma, and R.~Srikant.
\newblock Scaling up all pairs similarity search.
\newblock In {\em WWW}, pages 131--140, 2007.

\bibitem{Proc:Broder}
A.~Z. Broder.
\newblock On the resemblance and containment of documents.
\newblock In {\em the Compression and Complexity of Sequences}, pages 21--29,
  Positano, Italy, 1997.

\bibitem{Proc:Broder_STOC98}
A.~Z. Broder, M.~Charikar, A.~M. Frieze, and M.~Mitzenmacher.
\newblock Min-wise independent permutations.
\newblock In {\em STOC}, pages 327--336, Dallas, TX, 1998.

\bibitem{Proc:Buehrer_WSDM08}
G.~Buehrer and K.~Chellapilla.
\newblock A scalable pattern mining approach to web graph compression with
  communities.
\newblock In {\em WSDM}, pages 95--106, Stanford, CA, 2008.

\bibitem{Proc:Carter_STOC77}
J.~L. Carter and M.~N. Wegman.
\newblock Universal classes of hash functions.
\newblock In {\em STOC}, pages 106--112, 1977.

\bibitem{Sibyl}
T.~Chandra, E.~Ie, K.~Goldman, T.~L. Llinares, J.~McFadden, F.~Pereira,
  J.~Redstone, T.~Shaked, and Y.~Singer.
\newblock Sibyl: a system for large scale machine learning.

\bibitem{Proc:Charikar}
M.~S. Charikar.
\newblock Similarity estimation techniques from rounding algorithms.
\newblock In {\em STOC}, pages 380--388, Montreal, Quebec, Canada, 2002.

\bibitem{Proc:Chien_WWW05}
S.~Chien and N.~Immorlica.
\newblock Semantic similarity between search engine queries using temporal
  correlation.
\newblock In {\em WWW}, pages 2--11, 2005.

\bibitem{Proc:Chierichetti_KDD09}
F.~Chierichetti, R.~Kumar, S.~Lattanzi, M.~Mitzenmacher, A.~Panconesi, and
  P.~Raghavan.
\newblock On compressing social networks.
\newblock In {\em KDD}, pages 219--228, Paris, France, 2009.

\bibitem{Proc:Fetterly_WWW03}
D.~Fetterly, M.~Manasse, M.~Najork, and J.~L. Wiener.
\newblock A large-scale study of the evolution of web pages.
\newblock In {\em WWW}, pages 669--678, Budapest, Hungary, 2003.

\bibitem{Proc:Henzinger_SIGIR06}
M.~R. Henzinger.
\newblock Finding near-duplicate web pages: a large-scale evaluation of
  algorithms.
\newblock In {\em SIGIR}, pages 284--291, 2006.

\bibitem{Proc:Indyk_STOC98}
P.~Indyk and R.~Motwani.
\newblock Approximate nearest neighbors: Towards removing the curse of
  dimensionality.
\newblock In {\em STOC}, pages 604--613, Dallas, TX, 1998.

\bibitem{Proc:Li_Church_EMNLP}
P.~Li and K.~W. Church.
\newblock Using sketches to estimate associations.
\newblock In {\em HLT/EMNLP}, pages 708--715, Vancouver, BC, Canada, 2005.

\bibitem{Proc:Li_Church_Hastie_NIPS06}
P.~Li, K.~W. Church, and T.~J. Hastie.
\newblock Conditional random sampling: A sketch-based sampling technique for
  sparse data.
\newblock In {\em NIPS}, pages 873--880, Vancouver, BC, Canada, 2006.

\bibitem{Proc:Li_Konig_NIPS10}
P.~Li, A.~C. {K\"{o}nig}, and W.~Gui.
\newblock b-bit minwise hashing for estimating three-way similarities.
\newblock In {\em Advances in Neural Information Processing Systems},
  Vancouver, BC, 2010.

\bibitem{Proc:Li_Owen_Zhang_NIPS12}
P.~Li, A.~B. Owen, and C.-H. Zhang.
\newblock One permutation hashing.
\newblock In {\em NIPS}, Lake Tahoe, NV, 2012.

\bibitem{Proc:HashLearning_NIPS11}
P.~Li, A.~Shrivastava, J.~Moore, and A.~C. K\"onig.
\newblock Hashing algorithms for large-scale learning.
\newblock In {\em NIPS}, Granada, Spain, 2011.

\bibitem{Proc:Mitzenmacher_SODA08}
M.~Mitzenmacher and S.~Vadhan.
\newblock Why simple hash functions work: exploiting the entropy in a data
  stream.
\newblock In {\em SODA}, 2008.

\bibitem{Proc:Najork_WSDM09}
M.~Najork, S.~Gollapudi, and R.~Panigrahy.
\newblock Less is more: sampling the neighborhood graph makes salsa better and
  faster.
\newblock In {\em WSDM}, pages 242--251, Barcelona, Spain, 2009.

\bibitem{Proc:Nisan_STOC90}
N.~Nisan.
\newblock Pseudorandom generators for space-bounded computations.
\newblock In {\em Proceedings of the twenty-second annual ACM symposium on
  Theory of computing}, STOC, pages 204--212, 1990.

\bibitem{Proc:Shrivastava_NIPS13}
A.~Shrivastava and P.~Li.
\newblock Beyond pairwise: Provably fast algorithms for approximate k-way
  similarity search.
\newblock In {\em NIPS}, Lake Tahoe, NV, 2013.

\bibitem{Proc:OneHashLSH_ICML14}
A.~Shrivastava and P.~Li.
\newblock Densifying one permutation hashing via rotation for fast near
  neighbor search.
\newblock In {\em ICML}, Beijing, China, 2014.

\bibitem{GoogleBlog}
S.~Tong.
\newblock Lessons learned developing a practical large scale machine learning
  system.
\newblock
  http://googleresearch.blogspot.com/2010/04/lessons-learned-developing-practical.html,
  2008.

\bibitem{Proc:Weinberger_ICML2009}
K.~Weinberger, A.~Dasgupta, J.~Langford, A.~Smola, and J.~Attenberg.
\newblock Feature hashing for large scale multitask learning.
\newblock In {\em ICML}, pages 1113--1120, 2009.

\end{thebibliography}

\end{document}